\documentclass[5p,final,number,sort&compress]{elsarticle}

\usepackage{amsmath}
\usepackage{amsfonts,bbold}
\usepackage{amssymb}
\usepackage{graphicx}
\usepackage{subfigure}
\usepackage{natbib}

\usepackage[colorlinks=true,linkcolor=blue,urlcolor=black,citecolor=blue]{hyperref}

\begin{document}

\begin{frontmatter}
\title{Geometric signature of non-Markovian dynamics}

\author[1]{Da-Wei Luo}
\ead{dawei.luo@stevens.edu}

\author[1]{Ting Yu\corref{cor}}
\ead{Ting.Yu@stevens.edu}
\affiliation[1]{organization={Center for Quantum Science and Engineering and Department of Physics},
    addressline={Stevens Institute of Technology},
    city={Hoboken},
    state={New Jersey},
    postcode={07030},
    country={USA}}
\cortext[cor]{Corresponding author}

\date{\today}

\begin{abstract}
Non-Markovian effects in the dynamics of an open system are typically characterized by non-monotonic information flows from the system to its environment or by information backflows from the environment to the system. Using a two-level system (TLS) coupled to a dissipative single-mode cavity, we demonstrate that the geometric decoherence of the open quantum system can serve as a reliable indicator of non-Markovian dynamics. This geometric approach also reveals finer details of the dynamics, such as the specific time points when non-Markovian behavior emerges. In particular, we show that the divergence of the geometric decoherence factor of the TLS can serve as a sufficient condition for non-Markovian dynamics, and in certain cases, it can even be both a necessary and sufficient condition.
\end{abstract}

\end{frontmatter}

\section{Introduction}

Noisy quantum evolution in open systems is of interest across various research fields, including quantum computing, quantum decoherence, quantum metrology, quantum error correction, and quantum-classical transitions~\cite{Nielsen2000,Breuer2002,Breuer2012a,Rivas2010a,Vacchini2011a,Loughlin2023a, Lidar2013a,Knill2000a, Rebentrost2009a,Viola1999a,Paz1993a,Hu1992a}. The transition from Markovian to non-Markovian regimes is a crucial and complex issue because non-Markovian effects depend on multiple physical parameters, including system-environment couplings, the spectrum structure of the environment, external driving forces, and quantum measurement processes~\cite{Breuer2002,Breuer2012a,Vega2017a,Vasile2014a,Gambetta2002a,Gambetta2003a,Diosi2008a}.

One useful approach to characterize and quantify the non-Markovian features of open systems is known as non-Markovianity~\cite{Breuer2016a,Vega2017a,Milz2019a,Pollock2018a}. However, the computation of non-Markovianity in open system dynamics can be challenging, particularly when the system's dimension is large or the desired master equations are unavailable. So far, the determination of non-Markovian dynamics has primarily focused on characterizing the divisibility~\cite{Laine2010a, Rivas2010a, Milz2019a} of the dynamical map and the information flows between the system and its environment~\cite{Breuer2012a,Chrui2018a,Fanchini2014u,Li2018a,Lu2010a,Rivas2014a,Smirne2010a,Wissmann2015a,Zhong2013a}. Due to the complexity of the quantum bath, its information is generally not directly accessible. Therefore, it is desirable to find an alternative method to extract properties of the quantum bath from the system's dynamics alone. In the context of designing error-tolerant or error-correction protocols, detailed knowledge of the quantum bath is often essential, which is referred to as quantum noise spectroscopy~\cite{Ferrie2018a,Khan2024a,Paz-Silva2017a,Wang2021a}.

Another important quantity in quantum dynamics is the geometric phase, which is associated with the evolution of quantum systems~\cite{Berry1984,Berry1990a,F-Wilczek1989a}. In open quantum systems, the system-environment information is encoded in a time-dependent effective Hamiltonian, which is typically non-Hermitian, giving rise to a complex-valued geometric phase. The real part of this phase represents the ordinary geometric phase arising from dissipative evolution, while the imaginary part can be used to characterize the decoherence process due to environmental influence. Similar to the geometric phase, this decoherence factor is of a geometric nature~\cite{Luo2018i,Luo2019d}, and it is expected to record information about the environment's memory and the features of quantum dynamics.

Although the geometric phase and non-Markovian dynamics of open quantum systems have been extensively studied~\cite{Aitchison1992a,Bassi2006,Carollo2003,Cui2012a,F-Wilczek1989a,Garrison1988a,Luo2018i,Luo2019d,Carollo2003,Chen2010a,Huang2008a,Zhang2012a}, the relationship between the geometric phase and the onset of non-Markovian dynamics in the interaction between a quantum state and a non-Markovian environment remains unclear. Thus, it is of interest to investigate how these two memory effects are interrelated within the underlying quantum dissipative dynamics. We show that the geometric decoherence of a dissipative open quantum system can reveal signatures of the system's dynamics and explore the potential for using quantum trajectory-based geometric decoherence as a computationally efficient witness~\cite{Dariusz2014a,Rivas2010a,Bylicka2017a,Hall2014a} of the non-Markovianity of quantum system dynamics. The geometric phase in open quantum systems  is clearly a history-dependent quantity. As such, it retains memory effects from the dynamical process or the environmental memory capacity, as measured by the bath correlation function.

In this paper, we demonstrate that our geometric approach provides detailed insights into system dynamics. This approach not only predicts whether the system's dynamics are non-Markovian, but also identifies the specific time points when transitions from Markovian to non-Markovian behavior occur. These transitions can be interpreted as the moments when information backflows begin. We show that the divergence of the imaginary part of the complex-valued geometric phase serves as a reliable indicator for uncovering the non-Markovian nature of the process.

The paper is organized as follows: In Section~\ref{sec_geo_decay}, we introduce geometric decoherence based on a controllable non-Markovian environment. In Section~\ref{sec_geo_sig}, we demonstrate that geometric decoherence can serve as a sufficient condition for signaling the onset of non-Markovian dynamics. We conclude and offer some comments in Section~\ref{sec_outro}. Some mathematical details are provided in~\ref{sec_append}.

\section{Geometric decoherence in a hierarchical environment} \label{sec_geo_decay}
We consider a hierarchically structured environment and analytically study the interplay between the geometric phase and the non-Markovian dynamics of the system. One of great advantages of using this model is that the combined environment offers more tunable channels for the information flows as seen from the discussions below. Our model consists of a two-level system (TLS) coupled to a single-mode cavity, which is then embedded in a bosonic bath~\cite{Ma2014g}, the cavity and bosonic bath together form a composite environment for the TLS. The total Hamiltonian is then given by (setting $\hbar=1$)
\begin{align}
    H_{\rm tot} &= \frac{\omega}{2}\sigma_z+\kappa[\sigma^-a ^\dagger+\sigma^+ a] + \omega_c a ^\dagger a \nonumber \\
    &+ \sum_k g_k [a b_k ^\dagger + a ^\dagger b_k] + \sum_k \omega_k b_k ^\dagger b_k,
\end{align}
where $\sigma_z$ is the Pauli-z operator for the TLS, with frequency $\omega$, $\sigma^{+(-)}$ its raising/lowering operator, $a(a ^\dagger)$ the annihilation (creation) operator for the single-mode cavity mode with a frequency of $\omega_c$, and $b_k(b_k ^\dagger)$ the annihilation (creation) operator for the $k$th environment mode with frequency $\omega_k$. $\kappa$ denotes the atom-cavity coupling strength, and $g_k$ denotes the coupling strength between the cavity and the $k$th bath mode. From the standpoint of the TLS, it is experiencing a hierarchical environment: a direct environment induced by the single cavity mode, and indirectly, the bosonic bath through coupling to the cavity mode.  The model can display a rich dynamical behavior, where the dynamics of the TLS can be tuned to be either Markov or non-Markovian regardless of whether the bosonic bath is in the non-Markovian region or not, and displays an interesting phase-diagram for the Markov to non-Markovian crossover~\cite{Ma2014g}. Here, the dynamics of the TLS will be exactly and analytically derived, without any approximations. The term ``non-Markovian'' or ``Markov'' here would refer to whether the dynamics of the TLS shows memory effects, signified by a non-monotonic trace distance change~\cite{Breuer2012a,Rivas2014a} or a negative rate coefficient in the master equation's dissipation term~\cite{Rivas2014a,Piilo2009r,Wolf2008b,Breuer2016a}.

To probe the dynamical properties of this open quantum system, we utilize the complex-valued geometric phase~\cite{Aitchison1992a}, whose imaginary part can signify the geometric contribution in decoherence~\cite{Luo2019d, Fiete2003a, Dente2011a, Vatasescu2022a, Omnes2011a}.  To investigate the geometric phase of the open-system dynamics~\cite{Breuer2002}, we find that the non-Markovian quantum state diffusion (QSD) equations~\cite{Diosi1995a,Diosi1998,Yu1999a,Percival1998a,Zhao2012a,Strunz2004a,Yu2004a} is particularly convenient since it is formulated as a set of quantum trajectories governed by a stochastic Schr\"{o}dinger equation with a non-Hermitian effective Hamiltonian. For this hierarchical environment, we may use a dual-noise projection technique~\cite{Chen2020i} that projects the two baths separately onto their own coherent state basis. After rotating out the both baths $\omega_c a ^\dagger a + \sum_k \omega_k b_k ^\dagger b_k$, and let $z_t^*=-i z^* e^{i \omega_c t}$ and $w_t^*=-i\sum_k g_k^* w_k^* e^{i \omega_k t}$, then the quantum trajectories represented by the pure state  $|\psi_{z^*,w^*}(t) \rangle = \langle z,w|\psi_{\rm tot}(t) \rangle$ satisfy the following stochastic Schr{\"o}dinger equation,
\begin{align}
    &\partial_t |\psi_{z^*,w^*}(t) \rangle = \left[-i H_s + L z_t^* - L^\dagger \bar{O}_z(t) \right.\nonumber \\
    & \left. - i w_t^* \bar{O}_z(t) -i z_t^* \bar{O}_w(t) \right] |\psi_{z^*,w^*}(t) \rangle \label{eq_qsd2}
\end{align}
where $H_s=\omega \sigma_z/2$ is the system Hamiltonian, $L=\kappa \sigma^-$ is the system-bath coupling operator. The reduced density operator $\rho_s=\mathrm{tr}_E(\rho_{tot})$ for the TLS may then be obtained as an ensemble average of the trajectories over the noises $z^*, w^*$,  $\rho_s=\mathcal{M}[|\psi_{z^*,w^*}(t) \rangle \langle \psi_{z,w}(t) |]$. We can show (see Appendix) that the O-operators corresponding to the functional derivative with respect to the stochastic noises $z(w)$ can be formally given as
\begin{align}
    \bar{O}_{z}(t) = F_z(t) \sigma^-, \,
    \bar{O}_{w}(t) = F_w(t) \sigma^-,
\end{align}
where
$\bar{O}_{z}(t) = \int_0^tds \alpha_{z}(t,s) \frac{\delta}{\delta z_s^*}$,
$\bar{O}_{w}(t) = \int_0^tds \alpha_{w}(t,s) \frac{\delta}{\delta w_s^*}$,
with the correlation functions for the two noises given by
$\alpha_z(t,s)=e^{-i \omega_c (t-s)}$
$\alpha_w(t,s)=\gamma_w\Gamma_w\exp[-\gamma_w |t-s|-i\omega_0(t-s)]/2$, assuming an Ornstein-Uhlenbeck noise $w_t$~\cite{Diosi1995a,Diosi1998}. The parameter $\Gamma_w$ denotes the coupling strength between the cavity and the bath, and $\gamma_w$ denotes the memory effects of the bath, with a larger $\gamma_w$ signifying weaker memory effects. In the limit $\gamma_w \rightarrow \infty$, the spectrum of the bath modes would be flat, and we have a memory-less bath with a correlation function $\alpha_w(t,s) \propto \delta(t-s)$. It can be further shown that the influence of the $w$ noise may be folded into the $z$ noise alone. With the Schr\"{o}dinger-like quantum trajectory equation Eq.~\eqref{eq_qsd2}, we can calculate the complex geometric phase as a function of time~\cite{Aitchison1992a,Luo2019d,Cui2014a}. In what follows, we consider the resonant case $\omega_c=\omega_0=\omega$. The geometric phase is obtained as the total phase minus the dynamical phase, $\beta = \varphi_T - \varphi_d$. Here, we choose an initial state for the TLS as $[\cos(\theta/2), \sin (\theta/2)]^T$. The generalized complex-valued geometric phase, after the ensemble average, is given by (see Appendix)
\begin{align}
    \beta
    &= \frac{1}{2} \left[\cos \theta (\omega t+i \log[g(t)] ) -i \log \eta_\beta \right] \label{eq_gph}
\end{align}
where
\begin{align}
  \eta_\beta = \frac{g(t) (1+\cos \theta) + (1 -\cos \theta) e^{i \omega t}}{g(t) (1-\cos \theta)+(1+\cos \theta) e^{i \omega t}}
\end{align}
and
\begin{equation}
	g(t)=\exp\left[-\kappa\int_0^tds F_z(s)\right] \in R
\end{equation}
can be analytically derived (for details, see Appendix).

\section{Geometric signatures as a sufficient condition for signaling the onset of non-Markovian dynamics} \label{sec_geo_sig}

\begin{figure}
    \centering
    \includegraphics[width=.41\textwidth]{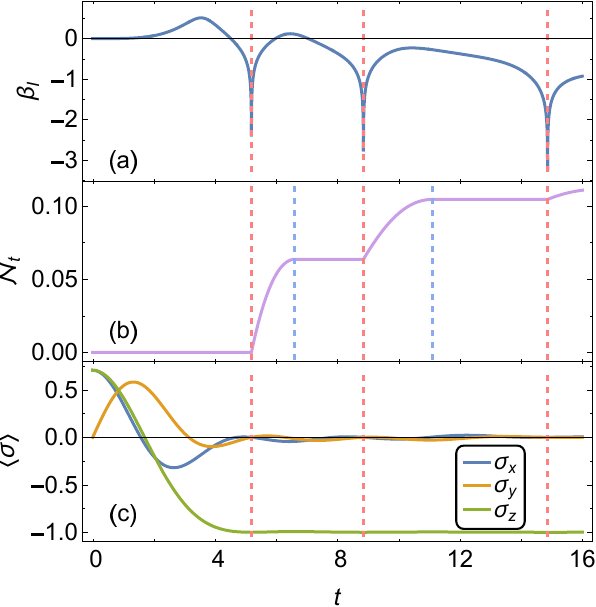}
    \caption{(Color online) Panel (a): The imaginary part of the complex-value geometric phase $\beta_I(t)$ as a function of time, in the parameter region where non-Markovian dynamics may be found ($\gamma_w=0.9$ and $\kappa = 0.43$). The dashed pink lines denote the time when $F_{z,R}(t)$ becomes negative and there is a back-flow of information from the bath to the system, which coincides with the time when the geometric decoherence divergence takes place at $t \approx 5.19, 8.85, 14.87$. Panel (b): Non-Markovianity as integrated up to time $t$, where it can be observed that the non-Markovianity measure increases in the region where the dynamics is non-contractive. Blue dashed lines denotes when $F_{z,R}(t)$ becomes positive and the dynamics becomes contractive. Panel (c): The expectation values $\langle\sigma_{x,y,z}\rangle$ as a function of time, which are smooth in time and do not display singular behaviors.} \label{fig_gpdiv}
\end{figure}

To quantify the degree of non-Markovianity, various useful measures have been proposed~\cite{Breuer2012a,Fanchini2014u,Rivas2014a, Breuer2016a,Shrikant2023a}. Among the approaches, some widely used are based on the notion of divisibility~\cite{Laine2010a, Rivas2010a, Milz2019a}, using dynamical semigroups~\cite{Davies1969a, Kossakowski1972a, Gorini1976a, Milnor1976a, Lindblad1976a, Hall2014a} and operational multi-time approaches~\cite{Pollock2018b, Pollock2018a, Budini2018a, Taranto2019a, Taranto2019b}. An intuitive picture is that the map representing the Markov dynamics is always contractive. Hence, the non-contractivity of the trace distance under a completely positive and trace-preserving map can be used to measure the non-Markovianity of an open quantum system~\cite{Breuer2012a,Laine2010a},
\begin{align}
    \mathcal{N} = \max_{\rho_1(0), \rho_2(0)}\int_{D_\sigma>0}dt D_\sigma(t, \rho_1(0), \rho_2(0)) \label{eq_nmdef}
\end{align}
where $D_\sigma(t, \rho_1, \rho_2) = D'(\rho_1(t),\rho_2(t))$ is the time derivative of the trace distance $D$ between the density operators evolved from a pair of initial states $(\rho_1(0), \rho_2(0))$ at time $t$. Here the trace distance between two quantum states $\rho_1, \rho_2$ is given by $D(\rho_1,\rho_2)=\mathrm{Tr}\sqrt{(\rho_1-\rho_2)^2}/2$. The integral in Eq.~\eqref{eq_nmdef} is optimized over all pairs of initial states $(\rho_1(0), \rho_2(0))$, which leads to $D(\rho_1(t),\rho_2(t)) = |g(t)|$~\cite{Ma2014g}.

It can be shown~\cite{Piilo2009r,Rivas2014a} (see Appendix) that when the real part of
\begin{equation}
	F_z(t) = -{g'(t)}/{\kappa g(t)}
\end{equation}
becomes negative, there will be information back-flow from the bath into the system, and the dynamical map is not contractive in these time ranges (denoted as $F_{z,R}(t) < 0$). Accordingly, the trace distance measure Eq.~\eqref{eq_nmdef} would begin to accumulate positive values.
We will take, without loss of generality, $\Gamma_w=1$. Choosing $\theta=\pi/4$, $\gamma_w=0.9$ and $\kappa = 0.43$, we plot the geometric decoherence factor given by the imaginary part of the geometric phase $\beta_I=\mathrm{Im}[\beta]$ as a function of time in Fig.~\ref{fig_gpdiv} (a), where the area with negative $F_{z,R}(t)$ signifies the time ranges where there is an information back-flow (or, equivalently, the dynamical map is not contractive), and the pink dashed line shows the boundary for these time ranges.
In Fig.~\ref{fig_gpdiv} (b), we plot the non-Markovianity measure Eq.~\eqref{eq_nmdef} integrated up to t,
\begin{align}
  \mathcal{N}_t = \max_{\rho_1(0), \rho_2(0)}\int_{D_\sigma>0, s=0}^{s=t}ds D_\sigma(s, \rho_1(0), \rho_2(0))
\end{align}
We can see that the non-Markovianity measure increases only in regions where the dynamics is non-contractive, otherwise it stays constant. This shows the agreement between the non-Markovianity measure based on the trace distance, and the non-contractive region as signified by the region where the coefficient of the dissipative terms in the master equation is negative, and the time points when the imaginary part of the geometric phase diverges.
It can also be seen that the geometric decoherence $\beta_I$ displays a sudden change and diverges at the boundary points, whereas the expectation values $\langle \sigma_{x,y,z} \rangle$ do not show any significant changes in Fig.~\ref{fig_gpdiv} (c). This shows that the detection of non-contractive dynamics can be realized with a pure geometric entity. The behavior of $\beta_I$ can be seen from the expression of the geometric phase Eq.~\eqref{eq_gph}: whenever the equation $g(t) = 0$ has roots for $t>0$, $\beta_I$ would diverge, and $F_z(t)$ would flip signs at the same time, an indication of the information back-flow. It is also worth pointing out that while the various non-Markovianity may not agree for arbitrary systems in general~\cite{Milz2019a, Breuer2016a,Rivas2014a,Shrikant2023a}, here the trace distance increases whenever the rate coefficient of the decoherence term in the master equation is negative. Since non-negative rate coefficient can be a necessary and sufficient condition for Markov or divisible processes~\cite{Rivas2014a,Wolf2008b,Breuer2016a}, for the model under consideration the divisible criteria and the trace distance measure agrees: the trace distance measure would be non-monotonic when the rate coefficient in the master equation's dissipative term become negative.

Moreover, the emergence of non-Markovian dynamical features may be associated with the concept of an information backflow from the bath into the system~\cite{Zhong2013a, Lu2010a, Vatasescu2022a}. The information flow may be quantified with the quantum Fisher information (QFI)~\cite{Helstrom1969a,Petz2011a} for a parameter $\theta$ in the density operator, defined as
\begin{equation}
	\mathcal{F}_\theta = \mathrm{tr}\left[\rho(\theta) \bar{L}_\theta^2\right], \label{eq_fi_def}
\end{equation}
with the symmetric logarithmic derivative operator $\bar{L}_\theta$ given by $\partial_\theta \rho(\theta) = \{ \rho(\theta), \bar{L}_\theta \} /2$.
With the parameter for the QFI being the angle $\theta$ in the initial state $|\psi(0) \rangle = \cos(\theta) |g \rangle + \sin(\theta)|e \rangle$ of the TLS, the evolved state of the TLS would be $\theta$-dependent. One can then calculate the associated QFI at any time $t$ for the TLS using~\eqref{eq_fi_def}, and it can be proven~\cite{Luo2024a} that there would be backflow of information from the bath back into the system whenever the dynamics is non-contractive. Thus, for the model under consideration, the non-Markovianity indicated by trace distance measures (non-contractivity), negative master equation rate term criteria (divisibility), and information flow measure using QFI are all in agreement.

\begin{figure}
    \centering
    \includegraphics[width=.41\textwidth]{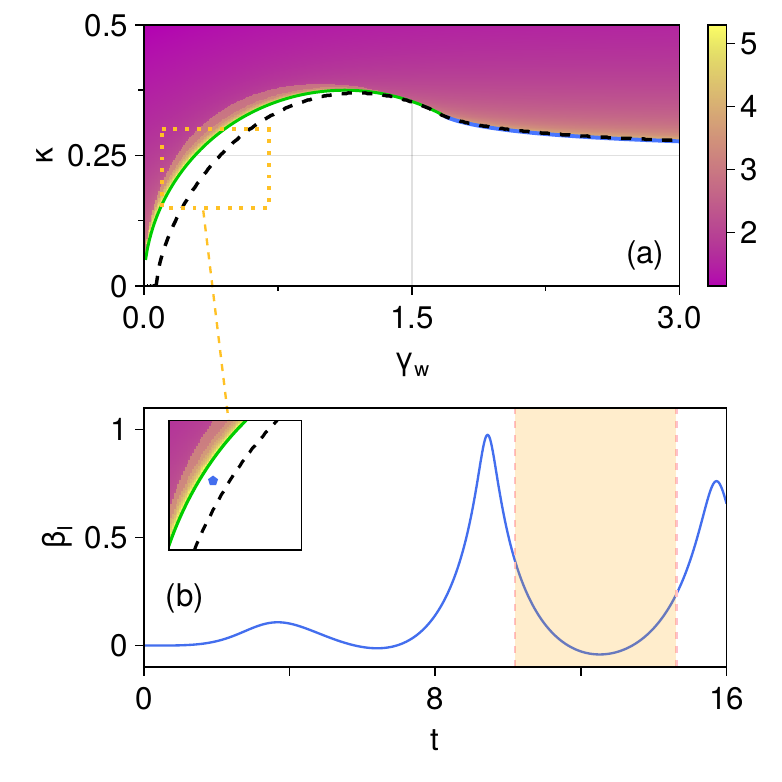}
    \caption{(Color online) Panel (a) Phase diagram shows when $\beta_I$ diverges, in the parameter domain ($\kappa,\gamma_w$). The color bar indicates the first divergent time point $t_0$ (log-scale). The black dashed line shows the boundary between the Markov and non-Markovian dynamics, where the region above the black dashed line is the non-Markovian region. However, when $\kappa, \gamma_\omega$ fall into the region enclosed by the black dashed line and the green line,  the system is still non-Markovian, but the divergence is not observed. Panel (b) Plot of $\beta_I(t)$ against time with the parameters $\gamma_w=0.3$ and $\kappa=0.23$, marked by a blue pentagon in the inset. Inset shows a zoom-in view of the area boxed by the dashed orange line in Panel (a).  Shaded area are time ranges when the information back-flow occurs.}
	\label{fig_pd}
\end{figure}

Our analysis suggests that a diverging $\beta_I$ for an arbitrary initial state parameter $\theta$ may serve as a sufficient condition for the TLS to experience non-Markovian dynamics. In Fig.~\ref{fig_pd}(a) we plot a phase diagram to show the parameter region and the earliest times when $\beta_I$ diverges. Numerical identification for divergence time points is carried over a time range of $[0,200]$. In fact, the boundary for whether the geometric phase can diverge, plotted as the green and blue lines in Fig.~\ref{fig_pd} may be analytically derived (see Appendix). The green boundary line is given by
\begin{align}
    \kappa = \frac{1}{6} \sqrt{\gamma_w (9-4 \gamma_w)},
\end{align}
and the blue boundary line satisfies
\begin{align}
    \kappa^2 &= \frac{1}{3} \sqrt{2 \gamma_w^3 \left(2 \gamma_w+27\right)} \cos \left(\frac{\theta}{3} \right)-\frac{1}{6} \gamma_w \left(4 \gamma_w+3\right), \nonumber \\
    \theta &= \sec ^{-1}\left(\frac{8 \sqrt{2 \gamma_w} \left(2 \gamma_w+27\right){}^{3/2}}{8 \gamma_w \left(4 \gamma_w-135\right)-729}\right).
\end{align}
The blue and green line joins at $\gamma_w=27/16, \kappa={3 \sqrt{3}}/{16}$.

Using Eq.~\eqref{eq_nmdef} over the same time range $t \in [0,200]$ as the search domain of diverging geometric decoherence factor $\beta_I$, we also calculate the non-Markovianity. The boundary of the crossover from Markov to non-Markovian dynamics is shown as the black dashed line in Fig.~\ref{fig_pd} (a), where non-Markovian dynamics is possible in the area above the black dashed line.
It can be seen that in the range $\gamma_w \in (0, 27/16]$ (boundary marked by in the green line), a diverging geometric decoherence for $\forall \theta$ may be a sufficient but not necessary condition for onset of non-Markovian dynamics for the TLS, while for $\gamma_w \in (27/16, 3]$ (boundary marked by the blue line) it may be a necessary and sufficient condition (numerically verified). A guess function for the Markov to non-Markovian boundary for $\gamma_w \in (0, 27/16]$ may also be analytically derived (see Appendix). This divergent behavior of the geometric decoherence suggests that it may be useful~\cite{Luo2014a,Ritz2000a,Cai2012a} for sensitive quantum metrology of small changes in the system parameter. Suppose one prepares the system near the transition lines in the phase diagram Fig~\ref{fig_pd}, then a small change in the system parameters can drive the dynamics of the TLS into a different region in the phase diagram, such that the presence or absence of sudden peaks in the geometric phase can serve as an indicator of whether a crossover has taken place. In contrast, as shown in Fig.~\ref{fig_gpdiv}, conventional measurement on the spin $\sigma_{x,y,z}$ may not show such a pronounced difference when the crossover takes place: the changes in the expectation values are less visible than the geometric decoherence factor $\beta_I$, so the latter may become more useful for the dynamics identification.

To understand this interplay between the geometric decoherence and Markov to non-Markovian crossover, we plot an example for non-diverging geometric phase in Fig.~\ref{fig_pd} (b) with $\gamma_w=0.3$ and $\kappa=0.23$, with orange-shaded area being time ranges when there is information backflow occurring. It can be seen that in this case, $\beta_I$ is non-divergent. For $\gamma_w \in (0, 27/16]$, the green line divides the non-Markovian region into two subregions: the one above the green line would show diverging geometric decoherence, whereas the region below the green line (but above the back dashed line) is a parameter region for ``exception'' cases where $\beta_I$ shows no divergence for non-Markovian dynamics. Mathematically, it is understood that in this region, we can have $g(t) > 0$ (i.e. geometric decoherence factor $\beta_I$ may be non-divergent) but with a non-monotonic $|g(t)|$ (i.e. TLS dynamics may be non-Markovian). One physical reason could be that in that region, while the indirect bath is technically non-Markovian, the coupling between the atom and its immediate bath are very weak in that region, so the influence of the bath may not be strong enough to cause the dramatic change in the geometric behavior of the atom.

\begin{figure}
    \centering
    \includegraphics[width=.41\textwidth]{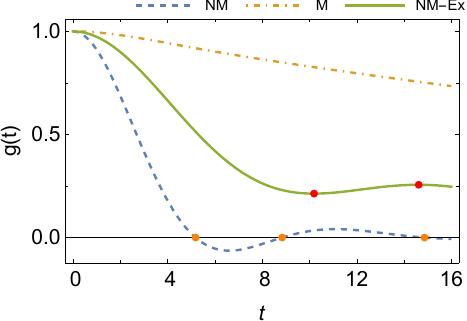}
    \caption{(Color online) The auxiliary function $g(t)$ as a function of time. The blue dashed line is from the non-Markovian region where the geometric phase can show a sudden jump, with the same parameters as Fig.~\ref{fig_gpdiv}. The orange dots on the line signifies times when $g(t)=0$ and $\partial_t|g(t)|$ becomes positive, such that the geometric phase diverges and the information flow between the system and bath changes direction. The green solid line is from the ``exceptional case'' where the system dynamics can be non-Markovian but the geometric phase does not diverge, with the same parameter as Fig.~\ref{fig_pd}(b), and the red dots shows the times $\partial_t g(t)$ flips signs. The orange dot-dashed line is from the Markovian region, with $\gamma_w=0.9$ and $\kappa=0.1$.}\label{fig_gt}
\end{figure}

In Fig.~\ref{fig_gt} we plot the auxiliary function $g(t)$ for the three different cases represented by a diverging geometric phase, and non-diverging geometric in both non-Markovian and Markov dynamics regions. It can be seen that for the case of divergent geometric decoherence, $g(t) = 0$ can have roots, and $\partial_t |g(t)|$ can become positive, indicative of a divergent geometric decoherence at the same time information flows from the bath back to the TLS (blue dashed line). Conversely, in the Markov region, we have $g(t) > 0$ and $g'(t)<0$, such that the geometric phase does not diverge, and the system dynamics shows no memory effects (orange dash-dotted line). In between those scenarios is the ``exceptional case'', where $g(t) >0$ but can be non-monotonic during certain time intervals (green solid line). In such cases, there is a backflow of information for times when $g(t)$ is increasing, but since $g(t) = 0$ has no roots, the geometric decoherence does not diverge. Mathematically, the relationship between a diverging geometric phase and non-Markovian dynamics is thus: when the equation $g(t) = 0$ has some root for $t > 0$, the geometric phase would diverge, while on the other hand, the existence of a root is a sufficient condition for $d|g(t)|/dt$ to become positive (or, equivalently, for $g'(t)$ and $g(t)$ to have the same sign), which indicates information back flow and the onset of non-Markovian dynamics.

The root $g(t)=0$ may be obtained by solving a transcendental equation, and in general, an analytical solution is not easily available. For our model,
we can have an explicit solution $g(t)=0$ in this limit $\gamma_w \rightarrow \infty$. It should be noted that the TLS can still experience the non-Markovian dynamics in this limit due to the hierarchical environment. We have~\cite{Ma2014g} (see Appendix)
\begin{align*}
    g(t) = e^{- \Gamma _w t /4 } \left[{\Gamma _w \sinh(c_M t/4)}/{c_M}+\cosh (c_M t/4)\right],
\end{align*}
where $c_M = \sqrt{\Gamma _w^2-16 \kappa ^2}$, and it can be shown that for $\kappa \leq \Gamma_w/4$, $g(t)>0$ and $g'(t)<0$, so the dynamics of the TLS stays Markov. For $\kappa > \Gamma_w/4$, $g(t)$ has roots (taking $\Gamma_w=1$, and $\kappa = 1/4 + \delta$ with $\delta > 0$),
\begin{align}
    t^{(n)}_{\mp} &= \frac{2 \sqrt{2} \left[n \pi \mp \tan ^{-1}(\varphi^{\pm} )\right]}{\sqrt{\delta (2 \delta +1)}}, \, \delta > 0
\end{align}
where $\varphi^+ = \sqrt{{2\delta } / {(2 \delta +1)} }$, $\varphi^- = 1/{\varphi^+}, \; n \in Z$. At such time points $t^{(n)}_{\mp}$ (only positive solutions are relevant), $\beta_I(t)$ diverges indicating the information back-flow occurs. Hence,
the system is non-Markovian. In this case, we have analytically shown that a diverging $\beta_I$ is a necessary and sufficient condition for the non-Markovian dynamics of the TLS.

\section{Concluding remarks} \label{sec_outro}

We have studied the relationship between non-Markovian dynamics and the geometric decoherence of a two level system coupled to a structured environment, consisting of a single mode cavity as directly-coupled environment, and an infinite mode bosonic bath coupled to the cavity mode as an indirect environment.
We find that the divergence of the geometric decoherence factor can indicate whether the dynamics of the TLS are non-Markovian, essentially serve as a strong signature of non-Markovian dynamics.
Specifically, the diverging geometric decoherence factor can serve as a strong sufficient condition for non-Markovian dynamics in the TLS. One intuitive explanation is that for open systems, the complex geometric phase contains information about both the phase and the amplitude (norm of the trajectories), which can be affected by non-Markovian effects. For example, changes in the amplitude can be correlated with changes in the trace distance, which quantifies non-Markovianity: under Markovian dynamics, the trace distance always decreases monotonically, whereas non-Markovian effects can cause the trace distance to increase. These turning points are dictated by the non-monotonic behavior of $|g(t)|$ which may cause singularities in the imaginary part of the generalized geometric phase.

Remarkably, under certain conditions, the divergent geometric phase constitutes both a necessary and sufficient criterion. Our findings reveal an intricate interplay between non-Markovian dynamics and the geometric decoherence factor, allowing one to extract properties of the quantum bath from the system's dynamics alone. We have shown that the geometric decoherence exhibits distinct behavior under non-Markovian dynamics~\cite{Luo2018i,Luo2019d} compared to memoryless baths. On the other hand, although the behaviors discussed here are model-specific, it would be very interesting to determine some more generic conditions under which an open system displays a diverging geometric decoherence signature during the Markov-to–non-Markovian crossover. The model discussed here is chosen for its rich physics and analytical solvability, allowing a detailed study of both the generalized complex geometric phase and non-Markovianity. Future investigations could extend this analysis to other models, such as a dephasing bath or the spin-Boson model.

We also note that the trace distance employed in the standard non-Markovianity measures are not a direct physical observable, and in more general cases a full state tomography may be needed. In addition, experimental measurements of non-Markovianity have been demonstrated for dephasing models~\cite{Li2020a,Tang2012a}, and it is plausible that such investigations could be extended to the model studied in this paper, which involves dissipative noise.

We would also like to point out that, for this model, we do not need to iterate over all possible initial states or optimize any function. In general, QSD trajectories work with $N$-dim pure state vectors rather than $N\times N$ density matrices,  which make the quantum trajectory approach computationally efficient. Therefore, this geometric signature can serve as a useful tool when resources for determining non-Markovian dynamics are limited or when rapid tracking of dynamical behaviors is crucial.  Finally, it would be of great interest to explore how geometric properties could be used to benchmark the degree of non-Markovianity for more general quantum systems.

\section{Acknowledgments}
This work is supported by the ART020-Quantum Technologies Project.

\appendix

\section{Derivation of the O-operator for the QSD} \label{sec_append}
Here, we derive the analytical expressions for the $\bar{O}_{z,w}$ operators for the QSD equation Eq.~\eqref{eq_qsd2} in the main text, with a similar procedure to~\cite{Chen2020i}.
For this model, we start with an ansatz $O_{z,w}(t,s)=f_{z,w}(t,s)\sigma^-$, with $F_{z,w}(t)=\int \alpha_{z,w}(t,s)f_{z,w}(t,s)ds$. Applying the consistence condition,
\begin{align}
    \partial_t \left[\frac{\delta}{\delta_{z,w}} |\psi_{z^*,w^*}(t) \rangle \right]= \frac{\delta}{\delta_{z,w}} \partial_t |\psi_{z^*,w^*}(t) \rangle
\end{align}
we have, since the $O_{z,w}$ operators are noise-free,
\begin{align}
    \dot{O}_{z,w}(t)
    & = \left[ -i H_s - L^\dagger \bar{O}_z(t), O_{z,w}(t) \right]
\end{align}
which gives,
\begin{align}
    \partial_t f_z(t,s) &= f_z(t,s)[i\omega+\kappa F_z(t)], \nonumber \\
    \partial_t f_w(t,s) &= f_w(t,s)[i\omega+\kappa F_z(t)],
\end{align}
with initial value condition given by ($L=\kappa \sigma^-$)
\begin{align}
    & O_z(t,t) =  L -i  \bar{O}_w(t) \nonumber \\
    & \Rightarrow f_z(t,t) = \kappa - iF_w(t)
\end{align}
and
\begin{align}
    & O_w(t,t) = -i \bar{O}_z(t) \nonumber \\
    & \Rightarrow f_w(t,t) = -i F_z(t)
\end{align}

Therefore, we have,
\begin{align}
    F_z'(t)
    &= \alpha_z(0)[\kappa - iF_w(t)] \nonumber \\
    & + i(\omega-\omega_c) F_z(t) +\kappa F_z^2(t),
\end{align}
where $'$ denotes derivative with respect to time, and
\begin{align}
    F_w'(t)
    &= -i\alpha_w(0)F_z(t) - \gamma_{w,\rm{eff}}F_w(t) \nonumber \\
    & + F_w(t,s)[i\omega+\kappa F_z(t)]
\end{align}
where $\gamma_{w,\rm{eff}} = \gamma_w + i \Omega_w$. Note we have set $\alpha_z(0)=1$ and $\alpha_w(0)=\gamma_w\Gamma_w/2$. The equations can be further simplified in the resonant case, $\omega=\omega_c=\Omega_w$,
\begin{align}
    F_z'(t)
    &= \kappa - iF_w(t) +\kappa F_z^2(t) \label{eq_dFz} \\
    F_w'(t)
    &= -i\gamma_w\Gamma_w F_z(t)/2 - \gamma_{w}F_w(t) + \kappa F_w(t) F_z(t) \label{eq_dFw}
\end{align}
The differential equation for $F_w$ may be further folded into a second order different equation for $F_z$,
\begin{align}
    F''_z(t) &= \frac{d}{dt} \left[\kappa - iF_w(t) +\kappa F_z^2(t)\right] \nonumber \\
    &= - iF'_w(t) + 2\kappa F_z(t)F'_z(t) \nonumber \\
    & \Rightarrow  F'_w(t) = i F''_z(t) - 2i \kappa F_z(t)F_z'(t) \label{eq_dFz2}
\end{align}
while from Eq.~\eqref{eq_dFz}, we have
\begin{align}
    F_w(t) = i [F'_z(t) - \kappa - \kappa F_z^2(t)]
\end{align}
Insert the $F_w$ and $F'_w$ into the RHS and LHS of Eq.~\eqref{eq_dFw}, we have
\begin{align}
    F''_z(t) &= \gamma_w \kappa -F_z'(t) [\gamma_w-3 \kappa F_z(t)] \nonumber \\
    & - \frac{1}{2} F_z(t) \left(\gamma_w \Gamma_w+ 2 \kappa  F_z(t) [\kappa  F_z(t)-\gamma_w]+2 \kappa ^2\right)
 \label{eq_dfz_t}
\end{align}
with boundary condition $F_z(0)=0$ and $F'_z(0) = \kappa$ from Eq.~\eqref{eq_dFz} and $F''_z(0) = 0$ from Eq.~\eqref{eq_dFz2}.

\section{Derivation of the master equation for the TLS}
The corresponding master equation may also be derived. Let $P_{z,w}= |\psi_{z^*,w^*}(t)\rangle\langle \psi_{z,w}(t)|$, we note that $z$ and $w$ are independent noises, and the $O_{z,w}$ operators here are noise-free, which lead to
\begin{align}
    \partial_t \rho_s &= \mathcal{M}_{z,w}[|\dot\psi_{z^*,w^*}\rangle\langle \psi_{z,w}|] + h.c. \nonumber \\
    &= -i [H_s, \rho_s] - L^\dagger \bar{O}_z(t) \rho_s + \bar{O}_z(t)\rho_s L^\dagger \nonumber \\
    & - \rho_s \bar{O}_z^\dagger(t) L + L\rho_s \bar{O}^\dagger_z(t),
\end{align}
where $\mathcal{M} \left[\cdot\right]$ is the ensemble average for the stochastic process,
and we have utilized the Novikov's theorem~\cite{Yu1999a}
\begin{align}
	\mathcal{M}_z[P_t z_t] &= \int_0^t ds \mathcal{M}[z_t z_s^*]\mathcal{M}[\frac{\delta P_t}{\delta z_s^*}] \nonumber \\
  & =\int_0^t ds \alpha(t,s)\mathcal{M}[\frac{\delta P_t}{\delta z_s^*}], \nonumber \\
	\mathcal{M}_z[P_t z_t^*] &= \int_0^t ds \mathcal{M}[z_t^* z_s]\mathcal{M}[\frac{\delta P_t}{\delta z_s}] \nonumber \\
  &= \int_0^t ds \alpha^*(t,s)\mathcal{M}[\frac{\delta P_t}{\delta z_s}].
\end{align}

Therefore, the master equation is formally only dependent on the $\bar{O}$ operator associated with the $z$-noise, (which is dependent on and includes the influences of the $\bar{O}$ operator associated with the $w$ noise)
\begin{align}
    \frac{\partial}{\partial t} \rho_s = -i \left[H_s,\rho_s\right]+\left[L,\rho_s \bar{O}_z^\dagger(t)\right]-\left[L^\dagger, \bar{O}_z(t) \rho_s\right]
\end{align}
With $\bar{O}_z(t)=F_z(t)\sigma^-\equiv F_{z,R} \sigma^- + i F_{z,I}\sigma^-$, and $L=\kappa \sigma^-$, it may be written in the Lindblad form. Define the Lindblad dissipator
\begin{align}
    \mathcal{D}[\sigma^-] = \sigma^- \rho_s \sigma^+
    -  \{ \sigma ^+ \sigma^-, \rho_s \}/2
\end{align}
we have
\begin{align}
    \frac{\partial}{\partial t} \rho_s &= -i \left[H_s+\kappa F_{z,I}(t) \sigma^+ \sigma^-,\rho_s\right] \nonumber \\
    & + 2 \kappa F_{z,R}(t) \mathcal{D}[\sigma^-],
\end{align}
where the dynamics may be considered non-contractive whenever $F_{z,R}(t)<0$.

\section{Derivation of the auxiliary g-function}
The equation~\eqref{eq_dfz_t} is a non-linear high order differential equation and may be difficult to analytically solve. However, if we introduce an auxiliary function
\begin{align}
    g(t)=\exp\left[-\kappa\int_0^tds F_z(s)\right],
\end{align}
which gives us,
\begin{align}
    F_z(t)& = -\frac{g'(t)}{\kappa g(t)}
\end{align}
continue with $g''(t)$ and $g'''(t)$ to get $\partial_t^n F_z$ in terms of $\partial_t^j g$
\begin{align}
    F_z'(t) &= \frac{\kappa ^2 g(t) F_z(t)^2-g''(t)}{\kappa  g(t)} \\
    F_z''(t) &= \frac{-g^{(3)}(t)+3 \kappa ^2 g(t) F_z(t) F_z'(t)-\kappa ^3 g(t) F_z(t)^3}{\kappa  g(t)}
\end{align}
and insert into Eq.~\eqref{eq_dfz_t}, we can have a differential equation for $g(t)$.
\begin{align}
    g'''(t) = -\gamma_w g''(t)-\frac{1}{2} \left(\gamma_w \Gamma_w+2 \kappa ^2\right) g'(t)-\gamma_w \kappa ^2 g(t) \label{dg_3}
\end{align}
with boundary condition $g(0)=1$, $g'(0)=0$, $g''(0) = -\kappa^2$ by using the boundary conditions for $\partial_t^n F_z(t)$.

Define a vector
\begin{align}
    v(t) = \begin{bmatrix}
        g(t)\\
        g'(t)\\
        g''(t)
    \end{bmatrix}
\end{align}
Eq.~\eqref{dg_3} may be written in a matrix form
\begin{align}
    \frac{d}{dt} v(t) = M v(t)
\end{align}
where
\begin{align}
    M = \left(
        \begin{array}{c|c|c}
         0 & 1 & 0 \\
         \hline
         0 & 0 & 1 \\
         \hline
         -\gamma_w \kappa ^2 & -\frac{1}{2} \left(\gamma_w \Gamma_w+2 \kappa ^2\right) & - \gamma_w \\
        \end{array}
        \right) \label{dg_mat}
\end{align}
where the first two rows are just identities for $g'$ and $g''$. The solution can then be written as
\begin{align}
    \exp(Mt)v(0).
\end{align}
It can now be shown,
\begin{align}
    g(t) = \sum_{x \in Rx} \frac{e^{\frac{t x}{2}} \left(2 \gamma _w \Gamma _w+2 x \gamma _w+x^2\right)}{4 \kappa ^2+2 \gamma _w \Gamma _w+4 x \gamma _w+3 x^2}
\end{align}
where $Rx = \{x_1,x_2,x_3\}$ are the three roots of the characteristic equation
\begin{align}
    x^3+2 x^2 \gamma _w+\left(4 \kappa ^2+2 \gamma _w \Gamma _w\right)x+8 \kappa ^2 \gamma _w = 0 \label{eq_chg}
\end{align}
It can be shown that this cubic equation has a determinant
\begin{align}
    D &\propto \gamma _w^2 \left(36 \kappa ^2-9 \gamma _w \Gamma _w+4 \gamma _w^2\right)^2 \nonumber \\
    & -2 \left(-6 \kappa ^2-3 \gamma _w \Gamma _w+2 \gamma _w^2\right)^3.\label{eq_gdet}
\end{align}
The cubic equation Eq.~\eqref{eq_chg} would have three real roots when $D\leq 0$, and when $D>0$ it would have one real root and a complex-conjugate pair as roots. In the latter case, $g(t)$ can be formally written as
\begin{align}
    g(t) = c_1 \exp(r_1 t) + \exp(r_2 t) \left[c_2 \sin(r_3 t) + c_3 \cos(r_3 t)\right],
\end{align}
where the coefficients $c(r)_{1\ldots3} \in R$ can be analytically derived. In both cases, we can show $g(t) \in R$.

\section{Derivation of the complex geometric phase}
Following a similar procedure to~\cite{Luo2019d} it can be straight-forwardly shown the total phase, after ensemble average, is given by
\begin{align}
    \varphi_{T}
    &= -i\log\left[\sqrt{g(t)}\right] \nonumber \\
    & -i \log\left[\sqrt{\frac{g(t) (1+\cos \theta) + (1 -\cos \theta) e^{i \omega t}}{g(t) (1-\cos \theta)+(1+\cos \theta) e^{i \omega t}}}\right]
\end{align}
Likewise, the dynamical phase
\begin{align}
    \varphi_d &= -i \int_0^tds \mathcal{M} \langle \tilde{\psi}(s)| \partial_s \psi(s) \rangle \nonumber \\
    &= -\int_0^tds \frac{1}{2} \left[\omega  \cos \theta -i \kappa  F_z(s) (\cos \theta +1) \right] \nonumber \\
    &= -\frac{1}{2} \omega t \cos \theta
    +\frac{i\kappa}{2} (\cos \theta +1) \int_0^t  F_z(s) ds
\end{align}
since $\log[g(t)]=-\kappa\int_0^tds F_z(s)$, the dynamical phase can be written, in terms of $g(t)$,
\begin{align}
    \varphi_d
    &= -\frac{1}{2} \omega t \cos \theta
    -\frac{i}{2} (\cos \theta +1) \log[g(t)]
\end{align}

\section{Derivation of the boundaries}

\begin{figure}
    \centering
    \subfigure[]{\includegraphics[scale=.7]{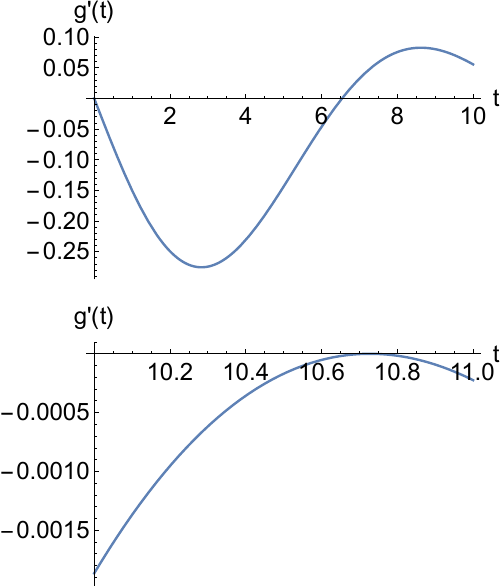}} \hspace{5mm}
    \subfigure[]{\includegraphics[scale=.45]{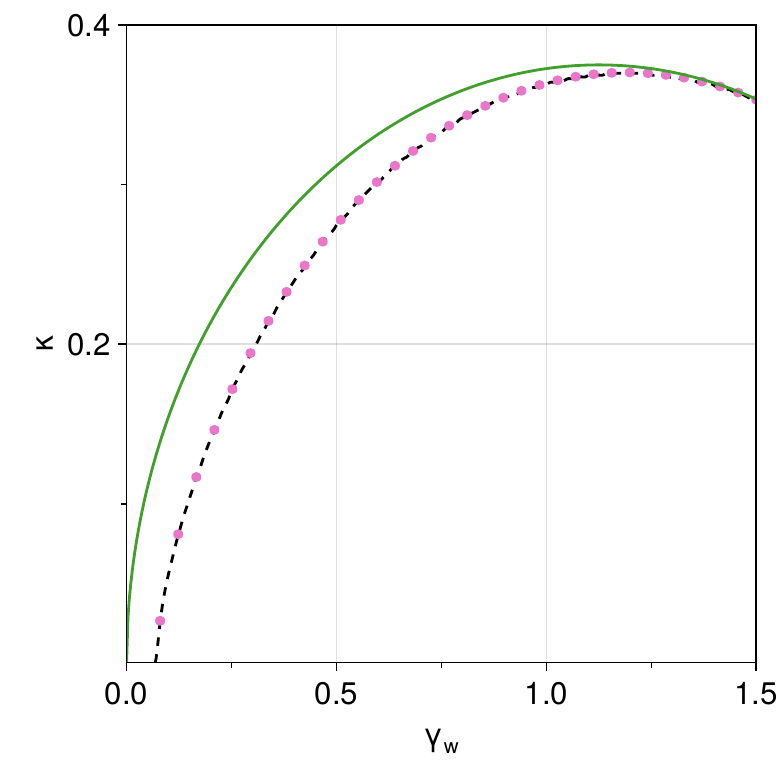}}
    \caption{(Color online) Panel (a) $g'(t)$ in the non-Markovian region $(\gamma_w =0.5, \kappa=0.4)$ in the upper plot, and on the boundary $(\gamma_w =0.5, \kappa \approx 0.27475)$ in the lower plot. It can be seen that $g'(t)$ intersects with the x-axis in the non-Markovian region but is tangent to the x-axis on the boundary. Panel (b) Analytically derived Markov to non-Markovian boundary for $\gamma_w \in [0.08, 1.5]$ shown as pink dots, where the black dashed line is the numerically obtained Markov to non-Markovian boundary, and the green line is the boundary between divergent and non-divergence geometric decoherence.}\label{fig_suppl_m2nm}
\end{figure}

Now we derive the trial boundaries for when the geometric phase can diverge, i.e. $g(t)=0$ can have roots. While $g(t)=0$ is a transcendental equation, we may use some numerical verification to guess the boundaries' analytical expressions. First note that the transcendental equation of the form
\begin{align}
    \exp(rt) = m \cos(a t + b)
\end{align}
is guaranteed to have roots $t>0$ when $r<0$ since the LHS would asymptotically approach $0$ for large $t$ while RHS oscillates between $[-m,m]$. On the other hand, when $rt>0$ and $|m|<1$ the equation is guaranteed to \emph{not} have roots since LHS would always be larger than $1$ while RHS would be smaller than $1$. Using this property, we may derive the green boundary line in Fig.~\ref{fig_pd} in the main text. We can also numerically verify in the region where the characteristic equation~\eqref{eq_chg} have all-real roots, $g(t)>0$ (no roots) and $g'(t)<0$ (always Markov dynamics). Thus, the blue boundary line in Fig.~\ref{fig_pd} in the main text may derived by setting the determinant~\eqref{eq_gdet} to zero.

In addition, for the range $\gamma_w \in (0,27/16)$, a reasonable guess for the roots to the equation $g'(t)=0$ may be that rather than being the intersection of $g'(t)$ with the x-axis, they would be tangents to the x-axis: i.e., we have $g'(t)=0$ and $g''(t)=0$ on the boundary. Solving for $g'(t)=0$ and $g''(t)=0$, we are also able to derive a guess function for the Markov to non-Markovian boundary for this region. In Fig.~\ref{fig_suppl_m2nm}, we show the different behaviors of the $g'(t)$ in the non-Markovian region and along the boundary, and the analytically derived Markov to non-Markovian boundary superimposed onto the numerically derived one, where an agreement between the two can be confirmed.

\section{Derivations in the Markov bath limit}
When the bath is in the memory-less limit $\gamma_w \rightarrow \infty$, it can be shown that non-Markovian dynamics of the TLS is a necessary and sufficient for the geometric phase to diverge. In the memory-less limit~\cite{Ma2014g},
\begin{align}
    g(t) = \mathcal{L}^{-1}\left[\frac{2 s+\Gamma_w}{2 \kappa ^2+2 s^2+s \Gamma_w}\right],
\end{align}
where $\mathcal{L}^{-1}$ is the inverse Laplace transform. When $\Gamma_w \neq 4 \kappa$
\begin{align}
    g(t) &= e^{-\Gamma _w t/4} \left(\frac{\Gamma _w \sinh \left(\frac{1}{4} t \sqrt{\Gamma _w^2-16 \kappa ^2}\right)}{\sqrt{\Gamma _w^2-16 \kappa ^2}} \right. \nonumber \\
    & \left. +\cosh \left(\frac{1}{4} t \sqrt{\Gamma _w^2-16 \kappa ^2}\right)\right),
\end{align}
whereas when $\Gamma_w = 4 \kappa$,
\begin{align}
    g(t) = \frac{1}{4} e^{-\Gamma _w t/4} \left(\Gamma _wt+4\right)
\end{align}

It can now be shown that for $\kappa \leq \Gamma_w/4$, $g(t)>0$ and $g'(t)<0$, while for $\kappa > \Gamma_w/4$, $g(t)$ has roots when, taking $\Gamma_w=1$, and let $\kappa = 1/4 + \delta$ with $\delta > 0$,
\begin{align}
    \frac{\sqrt{2} \sin \left(\frac{\sqrt{\delta  (2 \delta +1)} t}{\sqrt{2}}\right)}{\sqrt{\delta  (2 \delta +1)}}+4 \cos \left(\frac{\sqrt{\delta  (2 \delta +1)} t}{\sqrt{2}}\right) = 0,
\end{align}
i.e.
\begin{align}
    t &= \frac{2 \sqrt{2} \left(n \pi -\tan ^{-1}(\varphi )\right)}{\sqrt{\delta  (2 \delta +1)}}, \text{ or} \nonumber \\
    t &=  \frac{2 \sqrt{2} \left(n \pi +\tan ^{-1}\left(\frac{1}{\varphi }\right)\right)}{\sqrt{\delta  (2 \delta +1)}} \nonumber \\
    \varphi &= \frac{\sqrt{2} \delta }{\sqrt{\delta  (2 \delta +1)}}, \; n \in Z,
\end{align}
when the geometric phase would diverge at the same time the system dynamics becomes non-contractive and there is a flow of information from the bath back into the system.



\begin{thebibliography}{83}
\expandafter\ifx\csname natexlab\endcsname\relax\def\natexlab#1{#1}\fi
\providecommand{\url}[1]{\texttt{#1}}
\providecommand{\href}[2]{#2}
\providecommand{\path}[1]{#1}
\providecommand{\DOIprefix}{doi:}
\providecommand{\ArXivprefix}{arXiv:}
\providecommand{\URLprefix}{URL: }
\providecommand{\Pubmedprefix}{pmid:}
\providecommand{\doi}[1]{\href{http://dx.doi.org/#1}{\path{#1}}}
\providecommand{\Pubmed}[1]{\href{pmid:#1}{\path{#1}}}
\providecommand{\bibinfo}[2]{#2}
\ifx\xfnm\relax \def\xfnm[#1]{\unskip,\space#1}\fi
\bibitem[{Nielsen and Chuang(2000)}]{Nielsen2000}
\bibinfo{author}{M.~A. Nielsen}, \bibinfo{author}{I.~L. Chuang},
  \bibinfo{title}{{Quantum Computation and Quantum Information}},
  \bibinfo{publisher}{Cambridge University Press}, \bibinfo{year}{2000}.
\bibitem[{Breuer and Petruccione(2002)}]{Breuer2002}
\bibinfo{author}{H.~P. Breuer}, \bibinfo{author}{F.~Petruccione},
  \bibinfo{title}{{Theory of Open Quantum Systems}},
  \bibinfo{publisher}{Oxford, New York}, \bibinfo{year}{2002}.
\bibitem[{Breuer(2012)}]{Breuer2012a}
\bibinfo{author}{H.-P. Breuer},
\newblock \bibinfo{title}{{Foundations and measures of quantum
  non-Markovianity}},
\newblock \bibinfo{journal}{J. Phys. B} \bibinfo{volume}{45}
  (\bibinfo{year}{2012}) \bibinfo{pages}{154001}. \URLprefix
  \url{http://dx.doi.org/10.1088/0953-4075/45/15/154001}.
  \DOIprefix\doi{10.1088/0953-4075/45/15/154001}.
\bibitem[{Rivas et~al.(2010)Rivas, Huelga, and Plenio}]{Rivas2010a}
\bibinfo{author}{A.~Rivas}, \bibinfo{author}{S.~F. Huelga},
  \bibinfo{author}{M.~B. Plenio},
\newblock \bibinfo{title}{{Entanglement and Non-Markovianity of Quantum
  Evolutions}},
\newblock \bibinfo{journal}{Phys. Rev. Lett.} \bibinfo{volume}{105}
  (\bibinfo{year}{2010}) \bibinfo{pages}{050403}. \URLprefix
  \url{https://link.aps.org/doi/10.1103/PhysRevLett.105.050403}.
  \DOIprefix\doi{10.1103/PhysRevLett.105.050403}.
\bibitem[{Vacchini et~al.(2011)Vacchini, Smirne, Laine, Piilo, and
  Breuer}]{Vacchini2011a}
\bibinfo{author}{B.~Vacchini}, \bibinfo{author}{A.~Smirne},
  \bibinfo{author}{E.-M. Laine}, \bibinfo{author}{J.~Piilo},
  \bibinfo{author}{H.-P. Breuer},
\newblock \bibinfo{title}{{Markovianity and non-Markovianity in quantum and
  classical systems}},
\newblock \bibinfo{journal}{New Journal of Physics} \bibinfo{volume}{13}
  (\bibinfo{year}{2011}) \bibinfo{pages}{093004}. \URLprefix
  \url{http://dx.doi.org/10.1088/1367-2630/13/9/093004}.
  \DOIprefix\doi{10.1088/1367-2630/13/9/093004}.
\bibitem[{Loughlin and Sudhir(2023)}]{Loughlin2023a}
\bibinfo{author}{H.~A. Loughlin}, \bibinfo{author}{V.~Sudhir},
\newblock \bibinfo{title}{{Quantum noise and its evasion in feedback
  oscillators}},
\newblock \bibinfo{journal}{Nature Communications} \bibinfo{volume}{14}
  (\bibinfo{year}{2023}) \bibinfo{pages}{7083}. \URLprefix
  \url{https://doi.org/10.1038/s41467-023-42739-9}.
  \DOIprefix\doi{10.1038/s41467-023-42739-9}.
\bibitem[{Lidar and Brun(2013)}]{Lidar2013a}
\bibinfo{author}{D.~A. Lidar}, \bibinfo{author}{T.~A. Brun},
  \bibinfo{title}{{Quantum error correction}}, \bibinfo{publisher}{Cambridge
  university press}, \bibinfo{year}{2013}.
\bibitem[{Knill et~al.(2000)Knill, Laflamme, and Viola}]{Knill2000a}
\bibinfo{author}{E.~Knill}, \bibinfo{author}{R.~Laflamme},
  \bibinfo{author}{L.~Viola},
\newblock \bibinfo{title}{{Theory of Quantum Error Correction for General
  Noise}},
\newblock \bibinfo{journal}{Phys. Rev. Lett.} \bibinfo{volume}{84}
  (\bibinfo{year}{2000}) \bibinfo{pages}{2525--2528}. \URLprefix
  \url{https://link.aps.org/doi/10.1103/PhysRevLett.84.2525}.
  \DOIprefix\doi{10.1103/PhysRevLett.84.2525}.
\bibitem[{Rebentrost et~al.(2009)Rebentrost, Serban, Schulte-Herbr\"uggen, and
  Wilhelm}]{Rebentrost2009a}
\bibinfo{author}{P.~Rebentrost}, \bibinfo{author}{I.~Serban},
  \bibinfo{author}{T.~Schulte-Herbr\"uggen}, \bibinfo{author}{F.~K. Wilhelm},
\newblock \bibinfo{title}{{Optimal Control of a Qubit Coupled to a
  Non-Markovian Environment}},
\newblock \bibinfo{journal}{Phys. Rev. Lett.} \bibinfo{volume}{102}
  (\bibinfo{year}{2009}) \bibinfo{pages}{090401}. \URLprefix
  \url{https://link.aps.org/doi/10.1103/PhysRevLett.102.090401}.
  \DOIprefix\doi{10.1103/PhysRevLett.102.090401}.
\bibitem[{Viola et~al.(1999)Viola, Knill, and Lloyd}]{Viola1999a}
\bibinfo{author}{L.~Viola}, \bibinfo{author}{E.~Knill},
  \bibinfo{author}{S.~Lloyd},
\newblock \bibinfo{title}{{Dynamical Decoupling of Open Quantum Systems}},
\newblock \bibinfo{journal}{Phys. Rev. Lett.} \bibinfo{volume}{82}
  (\bibinfo{year}{1999}) \bibinfo{pages}{2417--2421}. \URLprefix
  \url{https://link.aps.org/doi/10.1103/PhysRevLett.82.2417}.
  \DOIprefix\doi{10.1103/PhysRevLett.82.2417}.
\bibitem[{Paz and Zurek(1993)}]{Paz1993a}
\bibinfo{author}{J.~P. Paz}, \bibinfo{author}{W.~H. Zurek},
\newblock \bibinfo{title}{{Environment-induced decoherence, classicality, and
  consistency of quantum histories}},
\newblock \bibinfo{journal}{Phys. Rev. D} \bibinfo{volume}{48}
  (\bibinfo{year}{1993}) \bibinfo{pages}{2728--2738}. \URLprefix
  \url{https://link.aps.org/doi/10.1103/PhysRevD.48.2728}.
  \DOIprefix\doi{10.1103/PhysRevD.48.2728}.
\bibitem[{Hu et~al.(1992)Hu, Paz, and Zhang}]{Hu1992a}
\bibinfo{author}{B.~L. Hu}, \bibinfo{author}{J.~P. Paz},
  \bibinfo{author}{Y.~Zhang},
\newblock \bibinfo{title}{{Quantum Brownian motion in a general environment:
  Exact master equation with nonlocal dissipation and colored noise}},
\newblock \bibinfo{journal}{Phys. Rev. D} \bibinfo{volume}{45}
  (\bibinfo{year}{1992}) \bibinfo{pages}{2843--2861}. \URLprefix
  \url{https://link.aps.org/doi/10.1103/PhysRevD.45.2843}.
  \DOIprefix\doi{10.1103/PhysRevD.45.2843}.
\bibitem[{de~Vega and Alonso(2017)}]{Vega2017a}
\bibinfo{author}{I.~de~Vega}, \bibinfo{author}{D.~Alonso},
\newblock \bibinfo{title}{{Dynamics of non-Markovian open quantum systems}},
\newblock \bibinfo{journal}{Rev. Mod. Phys.} \bibinfo{volume}{89}
  (\bibinfo{year}{2017}) \bibinfo{pages}{015001}. \URLprefix
  \url{https://link.aps.org/doi/10.1103/RevModPhys.89.015001}.
  \DOIprefix\doi{10.1103/RevModPhys.89.015001}.
\bibitem[{Vasile et~al.(2014)Vasile, Galve, and Zambrini}]{Vasile2014a}
\bibinfo{author}{R.~Vasile}, \bibinfo{author}{F.~Galve},
  \bibinfo{author}{R.~Zambrini},
\newblock \bibinfo{title}{{Spectral origin of non-Markovian open-system
  dynamics: A finite harmonic model without approximations}},
\newblock \bibinfo{journal}{Phys. Rev. A} \bibinfo{volume}{89}
  (\bibinfo{year}{2014}) \bibinfo{pages}{022109}. \URLprefix
  \url{https://link.aps.org/doi/10.1103/PhysRevA.89.022109}.
  \DOIprefix\doi{10.1103/PhysRevA.89.022109}.
\bibitem[{Gambetta and Wiseman(2002)}]{Gambetta2002a}
\bibinfo{author}{J.~Gambetta}, \bibinfo{author}{H.~M. Wiseman},
\newblock \bibinfo{title}{{Non-Markovian stochastic Schr\"odinger equations:
  Generalization to real-valued noise using quantum-measurement theory}},
\newblock \bibinfo{journal}{Phys. Rev. A} \bibinfo{volume}{66}
  (\bibinfo{year}{2002}) \bibinfo{pages}{012108}. \URLprefix
  \url{http://link.aps.org/doi/10.1103/PhysRevA.66.012108}.
  \DOIprefix\doi{10.1103/PhysRevA.66.012108}.
\bibitem[{Gambetta and Wiseman(2003)}]{Gambetta2003a}
\bibinfo{author}{J.~Gambetta}, \bibinfo{author}{H.~M. Wiseman},
\newblock \bibinfo{title}{{Interpretation of non-Markovian stochastic
  Schr\"odinger equations as a hidden-variable theory}},
\newblock \bibinfo{journal}{Phys. Rev. A} \bibinfo{volume}{68}
  (\bibinfo{year}{2003}) \bibinfo{pages}{062104}. \URLprefix
  \url{https://link.aps.org/doi/10.1103/PhysRevA.68.062104}.
  \DOIprefix\doi{10.1103/PhysRevA.68.062104}.
\bibitem[{Di\'osi(2008)}]{Diosi2008a}
\bibinfo{author}{L.~Di\'osi},
\newblock \bibinfo{title}{{Non-Markovian Continuous Quantum Measurement of
  Retarded Observables}},
\newblock \bibinfo{journal}{Phys. Rev. Lett.} \bibinfo{volume}{100}
  (\bibinfo{year}{2008}) \bibinfo{pages}{080401}. \URLprefix
  \url{https://link.aps.org/doi/10.1103/PhysRevLett.100.080401}.
  \DOIprefix\doi{10.1103/PhysRevLett.100.080401}.
\bibitem[{Breuer et~al.(2016)Breuer, Laine, Piilo, and Vacchini}]{Breuer2016a}
\bibinfo{author}{H.-P. Breuer}, \bibinfo{author}{E.-M. Laine},
  \bibinfo{author}{J.~Piilo}, \bibinfo{author}{B.~Vacchini},
\newblock \bibinfo{title}{{Colloquium: Non-Markovian dynamics in open quantum
  systems}},
\newblock \bibinfo{journal}{Rev. Mod. Phys.} \bibinfo{volume}{88}
  (\bibinfo{year}{2016}) \bibinfo{pages}{021002}. \URLprefix
  \url{https://link.aps.org/doi/10.1103/RevModPhys.88.021002}.
  \DOIprefix\doi{10.1103/RevModPhys.88.021002}.
\bibitem[{Milz et~al.(2019)Milz, Kim, Pollock, and Modi}]{Milz2019a}
\bibinfo{author}{S.~Milz}, \bibinfo{author}{M.~S. Kim}, \bibinfo{author}{F.~A.
  Pollock}, \bibinfo{author}{K.~Modi},
\newblock \bibinfo{title}{{Completely Positive Divisibility Does Not Mean
  Markovianity}},
\newblock \bibinfo{journal}{Phys. Rev. Lett.} \bibinfo{volume}{123}
  (\bibinfo{year}{2019}) \bibinfo{pages}{040401}. \URLprefix
  \url{https://link.aps.org/doi/10.1103/PhysRevLett.123.040401}.
  \DOIprefix\doi{10.1103/PhysRevLett.123.040401}.
\bibitem[{Pollock et~al.(2018)Pollock, Rodr\'{\i}guez-Rosario, Frauenheim,
  Paternostro, and Modi}]{Pollock2018a}
\bibinfo{author}{F.~A. Pollock}, \bibinfo{author}{C.~Rodr\'{\i}guez-Rosario},
  \bibinfo{author}{T.~Frauenheim}, \bibinfo{author}{M.~Paternostro},
  \bibinfo{author}{K.~Modi},
\newblock \bibinfo{title}{{Non-Markovian quantum processes: Complete framework
  and efficient characterization}},
\newblock \bibinfo{journal}{Phys. Rev. A} \bibinfo{volume}{97}
  (\bibinfo{year}{2018}) \bibinfo{pages}{012127}. \URLprefix
  \url{https://link.aps.org/doi/10.1103/PhysRevA.97.012127}.
  \DOIprefix\doi{10.1103/PhysRevA.97.012127}.
\bibitem[{Laine et~al.(2010)Laine, Piilo, and Breuer}]{Laine2010a}
\bibinfo{author}{E.-M. Laine}, \bibinfo{author}{J.~Piilo},
  \bibinfo{author}{H.-P. Breuer},
\newblock \bibinfo{title}{{Measure for the non-Markovianity of quantum
  processes}},
\newblock \bibinfo{journal}{Phys. Rev. A} \bibinfo{volume}{81}
  (\bibinfo{year}{2010}) \bibinfo{pages}{062115}. \URLprefix
  \url{https://link.aps.org/doi/10.1103/PhysRevA.81.062115}.
  \DOIprefix\doi{10.1103/PhysRevA.81.062115}.
\bibitem[{Chru\ifmmode \acute{s}\else \'{s}\fi{}ci\ifmmode~\acute{n}\else
  \'{n}\fi{}ski et~al.(2018)Chru\ifmmode \acute{s}\else
  \'{s}\fi{}ci\ifmmode~\acute{n}\else \'{n}\fi{}ski, Rivas, and
  St\o{}rmer}]{Chrui2018a}
\bibinfo{author}{D.~Chru\ifmmode \acute{s}\else
  \'{s}\fi{}ci\ifmmode~\acute{n}\else \'{n}\fi{}ski},
  \bibinfo{author}{A.~Rivas}, \bibinfo{author}{E.~St\o{}rmer},
\newblock \bibinfo{title}{{Divisibility and Information Flow Notions of Quantum
  Markovianity for Noninvertible Dynamical Maps}},
\newblock \bibinfo{journal}{Phys. Rev. Lett.} \bibinfo{volume}{121}
  (\bibinfo{year}{2018}) \bibinfo{pages}{080407}. \URLprefix
  \url{https://link.aps.org/doi/10.1103/PhysRevLett.121.080407}.
  \DOIprefix\doi{10.1103/PhysRevLett.121.080407}.
\bibitem[{Fanchini et~al.(2014)Fanchini, Karpat, \ifmmode~\mbox{\c{C}}\else
  \c{C}\fi{}akmak, Castelano, Aguilar, Far\'{\i}as, Walborn, Ribeiro, and
  de~Oliveira}]{Fanchini2014u}
\bibinfo{author}{F.~F. Fanchini}, \bibinfo{author}{G.~Karpat},
  \bibinfo{author}{B.~\ifmmode~\mbox{\c{C}}\else \c{C}\fi{}akmak},
  \bibinfo{author}{L.~K. Castelano}, \bibinfo{author}{G.~H. Aguilar},
  \bibinfo{author}{O.~J. Far\'{\i}as}, \bibinfo{author}{S.~P. Walborn},
  \bibinfo{author}{P.~H.~S. Ribeiro}, \bibinfo{author}{M.~C. de~Oliveira},
\newblock \bibinfo{title}{{Non-Markovianity through Accessible Information}},
\newblock \bibinfo{journal}{Phys. Rev. Lett.} \bibinfo{volume}{112}
  (\bibinfo{year}{2014}) \bibinfo{pages}{210402}. \URLprefix
  \url{https://link.aps.org/doi/10.1103/PhysRevLett.112.210402}.
  \DOIprefix\doi{10.1103/PhysRevLett.112.210402}.
\bibitem[{Li et~al.(2018)Li, Hall, and Wiseman}]{Li2018a}
\bibinfo{author}{L.~Li}, \bibinfo{author}{M.~J. Hall}, \bibinfo{author}{H.~M.
  Wiseman},
\newblock \bibinfo{title}{{Concepts of quantum non-Markovianity: A hierarchy}},
\newblock \bibinfo{journal}{Physics Reports} \bibinfo{volume}{759}
  (\bibinfo{year}{2018}) \bibinfo{pages}{1--51}.
\bibitem[{Lu et~al.(2010)Lu, Wang, and Sun}]{Lu2010a}
\bibinfo{author}{X.-M. Lu}, \bibinfo{author}{X.~Wang}, \bibinfo{author}{C.~P.
  Sun},
\newblock \bibinfo{title}{{Quantum Fisher information flow and non-Markovian
  processes of open systems}},
\newblock \bibinfo{journal}{Phys. Rev. A} \bibinfo{volume}{82}
  (\bibinfo{year}{2010}) \bibinfo{pages}{042103}. \URLprefix
  \url{https://link.aps.org/doi/10.1103/PhysRevA.82.042103}.
  \DOIprefix\doi{10.1103/PhysRevA.82.042103}.
\bibitem[{Rivas et~al.(2014)Rivas, Huelga, and Plenio}]{Rivas2014a}
\bibinfo{author}{A.~Rivas}, \bibinfo{author}{S.~F. Huelga},
  \bibinfo{author}{M.~B. Plenio},
\newblock \bibinfo{title}{{Quantum non-Markovianity: characterization,
  quantification and detection}},
\newblock \bibinfo{journal}{Reports on Progress in Physics}
  \bibinfo{volume}{77} (\bibinfo{year}{2014}) \bibinfo{pages}{094001}.
  \URLprefix \url{http://dx.doi.org/10.1088/0034-4885/77/9/094001}.
  \DOIprefix\doi{10.1088/0034-4885/77/9/094001}.
\bibitem[{Smirne et~al.(2010)Smirne, Breuer, Piilo, and Vacchini}]{Smirne2010a}
\bibinfo{author}{A.~Smirne}, \bibinfo{author}{H.-P. Breuer},
  \bibinfo{author}{J.~Piilo}, \bibinfo{author}{B.~Vacchini},
\newblock \bibinfo{title}{{Initial correlations in open-systems dynamics: The
  Jaynes-Cummings model}},
\newblock \bibinfo{journal}{Phys. Rev. A} \bibinfo{volume}{82}
  (\bibinfo{year}{2010}) \bibinfo{pages}{062114}. \URLprefix
  \url{https://link.aps.org/doi/10.1103/PhysRevA.82.062114}.
  \DOIprefix\doi{10.1103/PhysRevA.82.062114}.
\bibitem[{Wi\ss{}mann et~al.(2015)Wi\ss{}mann, Breuer, and
  Vacchini}]{Wissmann2015a}
\bibinfo{author}{S.~Wi\ss{}mann}, \bibinfo{author}{H.-P. Breuer},
  \bibinfo{author}{B.~Vacchini},
\newblock \bibinfo{title}{{Generalized trace-distance measure connecting
  quantum and classical non-Markovianity}},
\newblock \bibinfo{journal}{Phys. Rev. A} \bibinfo{volume}{92}
  (\bibinfo{year}{2015}) \bibinfo{pages}{042108}. \URLprefix
  \url{https://link.aps.org/doi/10.1103/PhysRevA.92.042108}.
  \DOIprefix\doi{10.1103/PhysRevA.92.042108}.
\bibitem[{Zhong et~al.(2013)Zhong, Sun, Ma, Wang, and Nori}]{Zhong2013a}
\bibinfo{author}{W.~Zhong}, \bibinfo{author}{Z.~Sun}, \bibinfo{author}{J.~Ma},
  \bibinfo{author}{X.~Wang}, \bibinfo{author}{F.~Nori},
\newblock \bibinfo{title}{{Fisher information under decoherence in Bloch
  representation}},
\newblock \bibinfo{journal}{Phys. Rev. A} \bibinfo{volume}{87}
  (\bibinfo{year}{2013}) \bibinfo{pages}{022337}. \URLprefix
  \url{https://link.aps.org/doi/10.1103/PhysRevA.87.022337}.
  \DOIprefix\doi{10.1103/PhysRevA.87.022337}.
\bibitem[{Ferrie et~al.(2018)Ferrie, Granade, Paz-Silva, and
  Wiseman}]{Ferrie2018a}
\bibinfo{author}{C.~Ferrie}, \bibinfo{author}{C.~Granade},
  \bibinfo{author}{G.~Paz-Silva}, \bibinfo{author}{H.~M. Wiseman},
\newblock \bibinfo{title}{{Bayesian quantum noise spectroscopy}},
\newblock \bibinfo{journal}{New Journal of Physics} \bibinfo{volume}{20}
  (\bibinfo{year}{2018}) \bibinfo{pages}{123005}. \URLprefix
  \url{https://dx.doi.org/10.1088/1367-2630/aaf207}.
  \DOIprefix\doi{10.1088/1367-2630/aaf207}.
\bibitem[{Khan et~al.(2024)Khan, Dong, Norris, and Viola}]{Khan2024a}
\bibinfo{author}{M.~Q. Khan}, \bibinfo{author}{W.~Dong}, \bibinfo{author}{L.~M.
  Norris}, \bibinfo{author}{L.~Viola}, \bibinfo{title}{{SPAM-Robust Multi-axis
  Quantum Noise Spectroscopy in Temporally Correlated Environments}},
  \bibinfo{year}{2024}. \href{http://arxiv.org/abs/2402.12361}{{\tt
  arXiv:2402.12361}}.
\bibitem[{Paz-Silva et~al.(2017)Paz-Silva, Norris, and Viola}]{Paz-Silva2017a}
\bibinfo{author}{G.~A. Paz-Silva}, \bibinfo{author}{L.~M. Norris},
  \bibinfo{author}{L.~Viola},
\newblock \bibinfo{title}{{Multiqubit spectroscopy of Gaussian quantum noise}},
\newblock \bibinfo{journal}{Phys. Rev. A} \bibinfo{volume}{95}
  (\bibinfo{year}{2017}) \bibinfo{pages}{022121}. \URLprefix
  \url{https://link.aps.org/doi/10.1103/PhysRevA.95.022121}.
  \DOIprefix\doi{10.1103/PhysRevA.95.022121}.
\bibitem[{Wang and Clerk(2021)}]{Wang2021a}
\bibinfo{author}{Y.-X. Wang}, \bibinfo{author}{A.~A. Clerk},
\newblock \bibinfo{title}{{Intrinsic and induced quantum quenches for enhancing
  qubit-based quantum noise spectroscopy}},
\newblock \bibinfo{journal}{Nature Communications} \bibinfo{volume}{12}
  (\bibinfo{year}{2021}) \bibinfo{pages}{6528}. \URLprefix
  \url{https://doi.org/10.1038/s41467-021-26868-7}.
  \DOIprefix\doi{10.1038/s41467-021-26868-7}.
\bibitem[{Berry(1984)}]{Berry1984}
\bibinfo{author}{M.~V. Berry},
\newblock \bibinfo{title}{{Quantal Phase Factors Accompanying Adiabatic
  Changes}},
\newblock \bibinfo{journal}{Proc. R. Soc. A} \bibinfo{volume}{392}
  (\bibinfo{year}{1984}) \bibinfo{pages}{45--57}.
  \DOIprefix\doi{10.1098/rspa.1984.0023}.
\bibitem[{Berry(1990)}]{Berry1990a}
\bibinfo{author}{M.~V. Berry},
\newblock \bibinfo{title}{{Geometric Amplitude Factors in Adiabatic Quantum
  Transitions}},
\newblock \bibinfo{journal}{Proc. R. Soc. A} \bibinfo{volume}{430}
  (\bibinfo{year}{1990}) \bibinfo{pages}{405--411}.
  \DOIprefix\doi{10.1098/rspa.1990.0096}.
\bibitem[{Wilczek and Shapere(1989)}]{F-Wilczek1989a}
\bibinfo{author}{F.~Wilczek}, \bibinfo{author}{A.~Shapere},
  \bibinfo{title}{{Geometric Phases in Physics}}, \bibinfo{publisher}{World
  Scientific}, \bibinfo{year}{1989}.
\bibitem[{Luo et~al.(2018)Luo, You, Lin, Wu, and Yu}]{Luo2018i}
\bibinfo{author}{D.-W. Luo}, \bibinfo{author}{J.~Q. You},
  \bibinfo{author}{H.-Q. Lin}, \bibinfo{author}{L.-A. Wu},
  \bibinfo{author}{T.~Yu},
\newblock \bibinfo{title}{{Memory-induced geometric phase in non-Markovian open
  systems}},
\newblock \bibinfo{journal}{Phys. Rev. A} \bibinfo{volume}{98}
  (\bibinfo{year}{2018}) \bibinfo{pages}{052117}. \URLprefix
  \url{https://link.aps.org/doi/10.1103/PhysRevA.98.052117}.
  \DOIprefix\doi{10.1103/PhysRevA.98.052117}.
\bibitem[{Luo et~al.(2019)Luo, Lin, You, Wu, Chatterjee, and Yu}]{Luo2019d}
\bibinfo{author}{D.-W. Luo}, \bibinfo{author}{H.-Q. Lin},
  \bibinfo{author}{J.~Q. You}, \bibinfo{author}{L.-A. Wu},
  \bibinfo{author}{R.~Chatterjee}, \bibinfo{author}{T.~Yu},
\newblock \bibinfo{title}{{Geometric decoherence in diffusive open quantum
  systems}},
\newblock \bibinfo{journal}{Phys. Rev. A} \bibinfo{volume}{100}
  (\bibinfo{year}{2019}) \bibinfo{pages}{062112}. \URLprefix
  \url{https://link.aps.org/doi/10.1103/PhysRevA.100.062112}.
  \DOIprefix\doi{10.1103/PhysRevA.100.062112}.
\bibitem[{Aitchison and Wanelik(1992)}]{Aitchison1992a}
\bibinfo{author}{I.~J.~R. Aitchison}, \bibinfo{author}{K.~Wanelik},
\newblock \bibinfo{title}{{On the Real and Complex Geometric Phases}},
\newblock \bibinfo{journal}{Proceedings of the Royal Society of London A:
  Mathematical, Physical and Engineering Sciences} \bibinfo{volume}{439}
  (\bibinfo{year}{1992}) \bibinfo{pages}{25--34}.
  \DOIprefix\doi{10.1098/rspa.1992.0131}.
\bibitem[{Bassi and Ippoliti(2006)}]{Bassi2006}
\bibinfo{author}{A.~Bassi}, \bibinfo{author}{E.~Ippoliti},
\newblock \bibinfo{title}{{Geometric phase for open quantum systems and
  stochastic unravelings}},
\newblock \bibinfo{journal}{Phys. Rev. A} \bibinfo{volume}{73}
  (\bibinfo{year}{2006}) \bibinfo{pages}{062104}. \URLprefix
  \url{http://link.aps.org/doi/10.1103/PhysRevA.73.062104}.
  \DOIprefix\doi{10.1103/PhysRevA.73.062104}.
\bibitem[{Carollo et~al.(2003)Carollo, Fuentes-Guridi, Santos, and
  Vedral}]{Carollo2003}
\bibinfo{author}{A.~Carollo}, \bibinfo{author}{I.~Fuentes-Guridi},
  \bibinfo{author}{M.~F. Santos}, \bibinfo{author}{V.~Vedral},
\newblock \bibinfo{title}{{Geometric Phase in Open Systems}},
\newblock \bibinfo{journal}{Phys. Rev. Lett.} \bibinfo{volume}{90}
  (\bibinfo{year}{2003}) \bibinfo{pages}{160402}. \URLprefix
  \url{http://link.aps.org/doi/10.1103/PhysRevLett.90.160402}.
  \DOIprefix\doi{10.1103/PhysRevLett.90.160402}.
\bibitem[{Cui and Zheng(2012)}]{Cui2012a}
\bibinfo{author}{X.-D. Cui}, \bibinfo{author}{Y.~Zheng},
\newblock \bibinfo{title}{{Geometric phases in non-Hermitian quantum
  mechanics}},
\newblock \bibinfo{journal}{Phys. Rev. A} \bibinfo{volume}{86}
  (\bibinfo{year}{2012}) \bibinfo{pages}{064104}. \URLprefix
  \url{https://link.aps.org/doi/10.1103/PhysRevA.86.064104}.
  \DOIprefix\doi{10.1103/PhysRevA.86.064104}.
\bibitem[{Garrison and Wright(1988)}]{Garrison1988a}
\bibinfo{author}{J.~Garrison}, \bibinfo{author}{E.~Wright},
\newblock \bibinfo{title}{{Complex geometrical phases for dissipative
  systems}},
\newblock \bibinfo{journal}{Physics Letters A} \bibinfo{volume}{128}
  (\bibinfo{year}{1988}) \bibinfo{pages}{177 -- 181}. \URLprefix
  \url{http://www.sciencedirect.com/science/article/pii/037596018890905X}.
  \DOIprefix\doi{http://dx.doi.org/10.1016/0375-9601(88)90905-X}.
\bibitem[{Chen et~al.(2010)Chen, An, Tong, Luo, and Oh}]{Chen2010a}
\bibinfo{author}{J.-J. Chen}, \bibinfo{author}{J.-H. An},
  \bibinfo{author}{Q.-J. Tong}, \bibinfo{author}{H.-G. Luo},
  \bibinfo{author}{C.~H. Oh},
\newblock \bibinfo{title}{{Non-Markovian effect on the geometric phase of a
  dissipative qubit}},
\newblock \bibinfo{journal}{Phys. Rev. A} \bibinfo{volume}{81}
  (\bibinfo{year}{2010}) \bibinfo{pages}{022120}. \URLprefix
  \url{https://link.aps.org/doi/10.1103/PhysRevA.81.022120}.
  \DOIprefix\doi{10.1103/PhysRevA.81.022120}.
\bibitem[{Huang and Yi(2008)}]{Huang2008a}
\bibinfo{author}{X.~L. Huang}, \bibinfo{author}{X.~X. Yi},
\newblock \bibinfo{title}{{Non-Markovian effects on the geometric phase}},
\newblock \bibinfo{journal}{EPL (Europhysics Letters)} \bibinfo{volume}{82}
  (\bibinfo{year}{2008}) \bibinfo{pages}{50001}. \URLprefix
  \url{http://dx.doi.org/10.1209/0295-5075/82/50001}.
  \DOIprefix\doi{10.1209/0295-5075/82/50001}.
\bibitem[{Zhang et~al.(2012)Zhang, Lo, Xiong, Tu, and Nori}]{Zhang2012a}
\bibinfo{author}{W.-M. Zhang}, \bibinfo{author}{P.-Y. Lo},
  \bibinfo{author}{H.-N. Xiong}, \bibinfo{author}{M.~W.-Y. Tu},
  \bibinfo{author}{F.~Nori},
\newblock \bibinfo{title}{{General Non-Markovian Dynamics of Open Quantum
  Systems}},
\newblock \bibinfo{journal}{Phys. Rev. Lett.} \bibinfo{volume}{109}
  (\bibinfo{year}{2012}) \bibinfo{pages}{170402}. \URLprefix
  \url{https://link.aps.org/doi/10.1103/PhysRevLett.109.170402}.
  \DOIprefix\doi{10.1103/PhysRevLett.109.170402}.
\bibitem[{Chru\ifmmode \acute{s}\else \'{s}\fi{}ci\ifmmode~\acute{n}\else
  \'{n}\fi{}ski and Maniscalco(2014)}]{Dariusz2014a}
\bibinfo{author}{D.~Chru\ifmmode \acute{s}\else
  \'{s}\fi{}ci\ifmmode~\acute{n}\else \'{n}\fi{}ski},
  \bibinfo{author}{S.~Maniscalco},
\newblock \bibinfo{title}{{Degree of Non-Markovianity of Quantum Evolution}},
\newblock \bibinfo{journal}{Phys. Rev. Lett.} \bibinfo{volume}{112}
  (\bibinfo{year}{2014}) \bibinfo{pages}{120404}. \URLprefix
  \url{https://link.aps.org/doi/10.1103/PhysRevLett.112.120404}.
  \DOIprefix\doi{10.1103/PhysRevLett.112.120404}.
\bibitem[{Bylicka et~al.(2017)Bylicka, Johansson, and Ac\'{\i}n}]{Bylicka2017a}
\bibinfo{author}{B.~Bylicka}, \bibinfo{author}{M.~Johansson},
  \bibinfo{author}{A.~Ac\'{\i}n},
\newblock \bibinfo{title}{{Constructive Method for Detecting the Information
  Backflow of Non-Markovian Dynamics}},
\newblock \bibinfo{journal}{Phys. Rev. Lett.} \bibinfo{volume}{118}
  (\bibinfo{year}{2017}) \bibinfo{pages}{120501}. \URLprefix
  \url{https://link.aps.org/doi/10.1103/PhysRevLett.118.120501}.
  \DOIprefix\doi{10.1103/PhysRevLett.118.120501}.
\bibitem[{Hall et~al.(2014)Hall, Cresser, Li, and Andersson}]{Hall2014a}
\bibinfo{author}{M.~J.~W. Hall}, \bibinfo{author}{J.~D. Cresser},
  \bibinfo{author}{L.~Li}, \bibinfo{author}{E.~Andersson},
\newblock \bibinfo{title}{{Canonical form of master equations and
  characterization of non-Markovianity}},
\newblock \bibinfo{journal}{Phys. Rev. A} \bibinfo{volume}{89}
  (\bibinfo{year}{2014}) \bibinfo{pages}{042120}. \URLprefix
  \url{https://link.aps.org/doi/10.1103/PhysRevA.89.042120}.
  \DOIprefix\doi{10.1103/PhysRevA.89.042120}.
\bibitem[{Ma et~al.(2014)Ma, Chen, Chen, Hedemann, and Yu}]{Ma2014g}
\bibinfo{author}{T.~Ma}, \bibinfo{author}{Y.~Chen}, \bibinfo{author}{T.~Chen},
  \bibinfo{author}{S.~R. Hedemann}, \bibinfo{author}{T.~Yu},
\newblock \bibinfo{title}{{Crossover between non-Markovian and Markovian
  dynamics induced by a hierarchical environment}},
\newblock \bibinfo{journal}{Phys. Rev. A} \bibinfo{volume}{90}
  (\bibinfo{year}{2014}) \bibinfo{pages}{042108}. \URLprefix
  \url{https://link.aps.org/doi/10.1103/PhysRevA.90.042108}.
  \DOIprefix\doi{10.1103/PhysRevA.90.042108}.
\bibitem[{Piilo et~al.(2009)Piilo, H\"ark\"onen, Maniscalco, and
  Suominen}]{Piilo2009r}
\bibinfo{author}{J.~Piilo}, \bibinfo{author}{K.~H\"ark\"onen},
  \bibinfo{author}{S.~Maniscalco}, \bibinfo{author}{K.-A. Suominen},
\newblock \bibinfo{title}{{Open system dynamics with non-Markovian quantum
  jumps}},
\newblock \bibinfo{journal}{Phys. Rev. A} \bibinfo{volume}{79}
  (\bibinfo{year}{2009}) \bibinfo{pages}{062112}. \URLprefix
  \url{https://link.aps.org/doi/10.1103/PhysRevA.79.062112}.
  \DOIprefix\doi{10.1103/PhysRevA.79.062112}.
\bibitem[{Wolf and Cirac(2008)}]{Wolf2008b}
\bibinfo{author}{M.~M. Wolf}, \bibinfo{author}{J.~I. Cirac},
\newblock \bibinfo{title}{Dividing quantum channels},
\newblock \bibinfo{journal}{Communications in Mathematical Physics}
  \bibinfo{volume}{279} (\bibinfo{year}{2008}) \bibinfo{pages}{147--168}.
  \URLprefix \url{https://doi.org/10.1007/s00220-008-0411-y}.
  \DOIprefix\doi{10.1007/s00220-008-0411-y}.
\bibitem[{Fiete and Heller(2003)}]{Fiete2003a}
\bibinfo{author}{G.~A. Fiete}, \bibinfo{author}{E.~J. Heller},
\newblock \bibinfo{title}{{Semiclassical theory of coherence and decoherence}},
\newblock \bibinfo{journal}{Phys. Rev. A} \bibinfo{volume}{68}
  (\bibinfo{year}{2003}) \bibinfo{pages}{022112}. \URLprefix
  \url{https://link.aps.org/doi/10.1103/PhysRevA.68.022112}.
  \DOIprefix\doi{10.1103/PhysRevA.68.022112}.
\bibitem[{Dente et~al.(2011)Dente, Zangara, and Pastawski}]{Dente2011a}
\bibinfo{author}{A.~D. Dente}, \bibinfo{author}{P.~R. Zangara},
  \bibinfo{author}{H.~M. Pastawski},
\newblock \bibinfo{title}{{Non-Markovian decay and dynamics of decoherence in
  private and public environments}},
\newblock \bibinfo{journal}{Phys. Rev. A} \bibinfo{volume}{84}
  (\bibinfo{year}{2011}) \bibinfo{pages}{042104}. \URLprefix
  \url{https://link.aps.org/doi/10.1103/PhysRevA.84.042104}.
  \DOIprefix\doi{10.1103/PhysRevA.84.042104}.
\bibitem[{Vatasescu(2022)}]{Vatasescu2022a}
\bibinfo{author}{M.~Vatasescu},
\newblock \bibinfo{title}{{Dynamics of quantum Fisher information from a
  time-local non-Markovian master equation with decoherence rates and operators
  depending on the estimated parameter}},
\newblock \bibinfo{journal}{Phys. Rev. A} \bibinfo{volume}{106}
  (\bibinfo{year}{2022}) \bibinfo{pages}{042204}. \URLprefix
  \url{https://link.aps.org/doi/10.1103/PhysRevA.106.042204}.
  \DOIprefix\doi{10.1103/PhysRevA.106.042204}.
\bibitem[{Omn{\`e}s(2011)}]{Omnes2011a}
\bibinfo{author}{R.~Omn{\`e}s},
\newblock \bibinfo{title}{{Decoherence and Wave Function Collapse}},
\newblock \bibinfo{journal}{Foundations of Physics} \bibinfo{volume}{41}
  (\bibinfo{year}{2011}) \bibinfo{pages}{1857--1880}. \URLprefix
  \url{https://doi.org/10.1007/s10701-011-9588-6}.
  \DOIprefix\doi{10.1007/s10701-011-9588-6}.
\bibitem[{Di\'osi et~al.(1995)Di\'osi, Gisin, Halliwell, and
  Percival}]{Diosi1995a}
\bibinfo{author}{L.~Di\'osi}, \bibinfo{author}{N.~Gisin},
  \bibinfo{author}{J.~Halliwell}, \bibinfo{author}{I.~C. Percival},
\newblock \bibinfo{title}{{Decoherent Histories and Quantum State Diffusion}},
\newblock \bibinfo{journal}{Phys. Rev. Lett.} \bibinfo{volume}{74}
  (\bibinfo{year}{1995}) \bibinfo{pages}{203--207}. \URLprefix
  \url{https://link.aps.org/doi/10.1103/PhysRevLett.74.203}.
  \DOIprefix\doi{10.1103/PhysRevLett.74.203}.
\bibitem[{Di\'osi et~al.(1998)Di\'osi, Gisin, and Strunz}]{Diosi1998}
\bibinfo{author}{L.~Di\'osi}, \bibinfo{author}{N.~Gisin},
  \bibinfo{author}{W.~T. Strunz},
\newblock \bibinfo{title}{{Non-Markovian quantum state diffusion}},
\newblock \bibinfo{journal}{Phys. Rev. A} \bibinfo{volume}{58}
  (\bibinfo{year}{1998}) \bibinfo{pages}{1699--1712}.
\bibitem[{Yu et~al.(1999)Yu, Di\'osi, Gisin, and Strunz}]{Yu1999a}
\bibinfo{author}{T.~Yu}, \bibinfo{author}{L.~Di\'osi},
  \bibinfo{author}{N.~Gisin}, \bibinfo{author}{W.~T. Strunz},
\newblock \bibinfo{title}{{Non-Markovian quantum-state diffusion: Perturbation
  approach}},
\newblock \bibinfo{journal}{Phys. Rev. A} \bibinfo{volume}{60}
  (\bibinfo{year}{1999}) \bibinfo{pages}{91--103}. \URLprefix
  \url{https://link.aps.org/doi/10.1103/PhysRevA.60.91}.
  \DOIprefix\doi{10.1103/PhysRevA.60.91}.
\bibitem[{Percival(1998)}]{Percival1998a}
\bibinfo{author}{I.~Percival}, \bibinfo{title}{{Quantum state diffusion}},
  \bibinfo{publisher}{Cambridge University Press}, \bibinfo{year}{1998}.
\bibitem[{Zhao et~al.(2012)Zhao, Shi, Wu, and Yu}]{Zhao2012a}
\bibinfo{author}{X.~Zhao}, \bibinfo{author}{W.~Shi}, \bibinfo{author}{L.-A.
  Wu}, \bibinfo{author}{T.~Yu},
\newblock \bibinfo{title}{{Fermionic stochastic Schr\"odinger equation and
  master equation: An open-system model}},
\newblock \bibinfo{journal}{Phys. Rev. A} \bibinfo{volume}{86}
  (\bibinfo{year}{2012}) \bibinfo{pages}{032116}. \URLprefix
  \url{https://link.aps.org/doi/10.1103/PhysRevA.86.032116}.
  \DOIprefix\doi{10.1103/PhysRevA.86.032116}.
\bibitem[{Strunz and Yu(2004)}]{Strunz2004a}
\bibinfo{author}{W.~T. Strunz}, \bibinfo{author}{T.~Yu},
\newblock \bibinfo{title}{{Convolutionless Non-Markovian master equations and
  quantum trajectories: Brownian motion}},
\newblock \bibinfo{journal}{Phys. Rev. A} \bibinfo{volume}{69}
  (\bibinfo{year}{2004}) \bibinfo{pages}{052115}. \URLprefix
  \url{https://link.aps.org/doi/10.1103/PhysRevA.69.052115}.
  \DOIprefix\doi{10.1103/PhysRevA.69.052115}.
\bibitem[{Yu(2004)}]{Yu2004a}
\bibinfo{author}{T.~Yu},
\newblock \bibinfo{title}{{Non-Markovian quantum trajectories versus master
  equations: Finite-temperature heat bath}},
\newblock \bibinfo{journal}{Phys. Rev. A} \bibinfo{volume}{69}
  (\bibinfo{year}{2004}) \bibinfo{pages}{062107}. \URLprefix
  \url{https://link.aps.org/doi/10.1103/PhysRevA.69.062107}.
  \DOIprefix\doi{10.1103/PhysRevA.69.062107}.
\bibitem[{Chen et~al.(2020)Chen, Ding, Shi, Jun, and Yu}]{Chen2020i}
\bibinfo{author}{Y.~Chen}, \bibinfo{author}{Q.~Ding}, \bibinfo{author}{W.~Shi},
  \bibinfo{author}{J.~Jun}, \bibinfo{author}{T.~Yu},
\newblock \bibinfo{title}{{Exact entanglement dynamics mediated by leaky
  optical cavities}},
\newblock \bibinfo{journal}{J. Phys. B} \bibinfo{volume}{53}
  (\bibinfo{year}{2020}) \bibinfo{pages}{125501}.
\bibitem[{Cui and Zheng(2014)}]{Cui2014a}
\bibinfo{author}{X.-D. Cui}, \bibinfo{author}{Y.~Zheng},
\newblock \bibinfo{title}{{Unification of the family of Garrison-Wright's
  phases}},
\newblock \bibinfo{journal}{Sci. Rep.} \bibinfo{volume}{4}
  (\bibinfo{year}{2014}) \bibinfo{pages}{5813 EP --}. \URLprefix
  \url{http://dx.doi.org/10.1038/srep05813}.
\bibitem[{Shrikant and Mandayam(2023)}]{Shrikant2023a}
\bibinfo{author}{U.~Shrikant}, \bibinfo{author}{P.~Mandayam},
\newblock \bibinfo{title}{{Quantum non-Markovianity: Overview and recent
  developments}},
\newblock \bibinfo{journal}{Frontiers in Quantum Science and Technology}
  \bibinfo{volume}{2} (\bibinfo{year}{2023}). \URLprefix
  \url{https://www.frontiersin.org/articles/10.3389/frqst.2023.1134583}.
  \DOIprefix\doi{10.3389/frqst.2023.1134583}.
\bibitem[{Davies(1969)}]{Davies1969a}
\bibinfo{author}{E.~B. Davies},
\newblock \bibinfo{title}{{Quantum stochastic processes}},
\newblock \bibinfo{journal}{Communications in Mathematical Physics}
  \bibinfo{volume}{15} (\bibinfo{year}{1969}) \bibinfo{pages}{277--304}.
  \URLprefix \url{https://doi.org/10.1007/BF01645529}.
  \DOIprefix\doi{10.1007/BF01645529}.
\bibitem[{Kossakowski(1972)}]{Kossakowski1972a}
\bibinfo{author}{A.~Kossakowski},
\newblock \bibinfo{title}{{On quantum statistical mechanics of non-Hamiltonian
  systems}},
\newblock \bibinfo{journal}{Reports on Mathematical Physics}
  \bibinfo{volume}{3} (\bibinfo{year}{1972}) \bibinfo{pages}{247--274}.
  \URLprefix
  \url{https://www.sciencedirect.com/science/article/pii/0034487772900109}.
  \DOIprefix\doi{https://doi.org/10.1016/0034-4877(72)90010-9}.
\bibitem[{Gorini et~al.(1976)Gorini, Kossakowski, and Sudarshan}]{Gorini1976a}
\bibinfo{author}{V.~Gorini}, \bibinfo{author}{A.~Kossakowski},
  \bibinfo{author}{E.~C.~G. Sudarshan},
\newblock \bibinfo{title}{{Completely positive dynamical semigroups of N-level
  systems}},
\newblock \bibinfo{journal}{Journal of Mathematical Physics}
  \bibinfo{volume}{17} (\bibinfo{year}{1976}) \bibinfo{pages}{821--825}.
  \URLprefix \url{https://doi.org/10.1063/1.522979}.
  \DOIprefix\doi{10.1063/1.522979}.
\bibitem[{Milnor(1976)}]{Milnor1976a}
\bibinfo{author}{J.~Milnor},
\newblock \bibinfo{title}{{Curvatures of left invariant metrics on lie
  groups}},
\newblock \bibinfo{journal}{Advances in Mathematics} \bibinfo{volume}{21}
  (\bibinfo{year}{1976}) \bibinfo{pages}{293--329}. \URLprefix
  \url{https://www.sciencedirect.com/science/article/pii/S0001870876800023}.
  \DOIprefix\doi{https://doi.org/10.1016/S0001-8708(76)80002-3}.
\bibitem[{Lindblad(1976)}]{Lindblad1976a}
\bibinfo{author}{G.~Lindblad},
\newblock \bibinfo{title}{{On the generators of quantum dynamical semigroups}},
\newblock \bibinfo{journal}{Comm. Math. Phys.} \bibinfo{volume}{48}
  (\bibinfo{year}{1976}) \bibinfo{pages}{119--130}.
\bibitem[{Pollock et~al.(2018)Pollock, Rodr\'{\i}guez-Rosario, Frauenheim,
  Paternostro, and Modi}]{Pollock2018b}
\bibinfo{author}{F.~A. Pollock}, \bibinfo{author}{C.~Rodr\'{\i}guez-Rosario},
  \bibinfo{author}{T.~Frauenheim}, \bibinfo{author}{M.~Paternostro},
  \bibinfo{author}{K.~Modi},
\newblock \bibinfo{title}{{Operational Markov Condition for Quantum
  Processes}},
\newblock \bibinfo{journal}{Phys. Rev. Lett.} \bibinfo{volume}{120}
  (\bibinfo{year}{2018}) \bibinfo{pages}{040405}. \URLprefix
  \url{https://link.aps.org/doi/10.1103/PhysRevLett.120.040405}.
  \DOIprefix\doi{10.1103/PhysRevLett.120.040405}.
\bibitem[{Budini(2018)}]{Budini2018a}
\bibinfo{author}{A.~A. Budini},
\newblock \bibinfo{title}{{Quantum Non-Markovian Processes Break Conditional
  Past-Future Independence}},
\newblock \bibinfo{journal}{Phys. Rev. Lett.} \bibinfo{volume}{121}
  (\bibinfo{year}{2018}) \bibinfo{pages}{240401}. \URLprefix
  \url{https://link.aps.org/doi/10.1103/PhysRevLett.121.240401}.
  \DOIprefix\doi{10.1103/PhysRevLett.121.240401}.
\bibitem[{Taranto et~al.(2019{\natexlab{a}})Taranto, Pollock, Milz, Tomamichel,
  and Modi}]{Taranto2019a}
\bibinfo{author}{P.~Taranto}, \bibinfo{author}{F.~A. Pollock},
  \bibinfo{author}{S.~Milz}, \bibinfo{author}{M.~Tomamichel},
  \bibinfo{author}{K.~Modi},
\newblock \bibinfo{title}{{Quantum Markov Order}},
\newblock \bibinfo{journal}{Phys. Rev. Lett.} \bibinfo{volume}{122}
  (\bibinfo{year}{2019}{\natexlab{a}}) \bibinfo{pages}{140401}. \URLprefix
  \url{https://link.aps.org/doi/10.1103/PhysRevLett.122.140401}.
  \DOIprefix\doi{10.1103/PhysRevLett.122.140401}.
\bibitem[{Taranto et~al.(2019{\natexlab{b}})Taranto, Milz, Pollock, and
  Modi}]{Taranto2019b}
\bibinfo{author}{P.~Taranto}, \bibinfo{author}{S.~Milz}, \bibinfo{author}{F.~A.
  Pollock}, \bibinfo{author}{K.~Modi},
\newblock \bibinfo{title}{{Structure of quantum stochastic processes with
  finite Markov order}},
\newblock \bibinfo{journal}{Phys. Rev. A} \bibinfo{volume}{99}
  (\bibinfo{year}{2019}{\natexlab{b}}) \bibinfo{pages}{042108}. \URLprefix
  \url{https://link.aps.org/doi/10.1103/PhysRevA.99.042108}.
  \DOIprefix\doi{10.1103/PhysRevA.99.042108}.
\bibitem[{Helstrom(1969)}]{Helstrom1969a}
\bibinfo{author}{C.~W. Helstrom},
\newblock \bibinfo{title}{{Quantum detection and estimation theory}},
\newblock \bibinfo{journal}{Journal of Statistical Physics} \bibinfo{volume}{1}
  (\bibinfo{year}{1969}) \bibinfo{pages}{231--252}. \URLprefix
  \url{https://doi.org/10.1007/BF01007479}. \DOIprefix\doi{10.1007/BF01007479}.
\bibitem[{Petz and Ghinea(2011)}]{Petz2011a}
\bibinfo{author}{D.~Petz}, \bibinfo{author}{C.~Ghinea},
\newblock \bibinfo{title}{{Introduction to quantum Fisher information}},
\newblock in: \bibinfo{booktitle}{{Quantum probability and related topics}},
  \bibinfo{publisher}{World Scientific}, \bibinfo{year}{2011}, pp.
  \bibinfo{pages}{261--281}.
\bibitem[{Luo and Yu(2024)}]{Luo2024a}
\bibinfo{author}{D.-W. Luo}, \bibinfo{author}{T.~Yu},
  \bibinfo{title}{{Jaynes-Cummings atoms coupled to a structured environment:
  Leakage elimination operators and the Petz recovery maps}},
  \bibinfo{year}{2024}. \href{http://arxiv.org/abs/2404.13762}{{\tt
  arXiv:2404.13762}}.
\bibitem[{Luo and Xu(2014)}]{Luo2014a}
\bibinfo{author}{D.-W. Luo}, \bibinfo{author}{J.-B. Xu},
\newblock \bibinfo{title}{{Sensitive chemical compass and quantum criticality
  at finite temperature}},
\newblock \bibinfo{journal}{Physics Letters A} \bibinfo{volume}{378}
  (\bibinfo{year}{2014}) \bibinfo{pages}{1481--1486}.
\bibitem[{Ritz et~al.(2000)Ritz, Adem, and Schulten}]{Ritz2000a}
\bibinfo{author}{T.~Ritz}, \bibinfo{author}{S.~Adem},
  \bibinfo{author}{K.~Schulten},
\newblock \bibinfo{title}{{A model for photoreceptor-based magnetoreception in
  birds}},
\newblock \bibinfo{journal}{Biophysical journal} \bibinfo{volume}{78}
  (\bibinfo{year}{2000}) \bibinfo{pages}{707--718}.
\bibitem[{Cai et~al.(2012)Cai, Ai, Quan, and Sun}]{Cai2012a}
\bibinfo{author}{C.~Y. Cai}, \bibinfo{author}{Q.~Ai}, \bibinfo{author}{H.~T.
  Quan}, \bibinfo{author}{C.~P. Sun},
\newblock \bibinfo{title}{{Sensitive chemical compass assisted by quantum
  criticality}},
\newblock \bibinfo{journal}{Phys. Rev. A} \bibinfo{volume}{85}
  (\bibinfo{year}{2012}) \bibinfo{pages}{022315}. \URLprefix
  \url{https://link.aps.org/doi/10.1103/PhysRevA.85.022315}.
  \DOIprefix\doi{10.1103/PhysRevA.85.022315}.
\bibitem[{Li et~al.(2020)Li, Guo, and Piilo}]{Li2020a}
\bibinfo{author}{C.-F. Li}, \bibinfo{author}{G.-C. Guo},
  \bibinfo{author}{J.~Piilo},
\newblock \bibinfo{title}{{Non-Markovian quantum dynamics: What is it good
  for?}},
\newblock \bibinfo{journal}{EPL (Europhysics Letters)} \bibinfo{volume}{128}
  (\bibinfo{year}{2020}) \bibinfo{pages}{30001}.
  \DOIprefix\doi{10.1209/0295-5075/128/30001}.
\bibitem[{Tang et~al.(2012)Tang, Li, Li, Zou, Guo, Breuer, Laine, and
  Piilo}]{Tang2012a}
\bibinfo{author}{J.-S. Tang}, \bibinfo{author}{C.-F. Li},
  \bibinfo{author}{Y.-L. Li}, \bibinfo{author}{X.-B. Zou},
  \bibinfo{author}{G.-C. Guo}, \bibinfo{author}{H.-P. Breuer},
  \bibinfo{author}{E.-M. Laine}, \bibinfo{author}{J.~Piilo},
\newblock \bibinfo{title}{{Measuring non-Markovianity of processes with
  controllable system-environment interaction}},
\newblock \bibinfo{journal}{EPL (Europhysics Letters)} \bibinfo{volume}{97}
  (\bibinfo{year}{2012}) \bibinfo{pages}{10002}.
  \DOIprefix\doi{10.1209/0295-5075/97/10002}.

\end{thebibliography}



\end{document}